\documentclass[fleqn,useAMS,usegraphicx,usedcolumn,usenatbib]{mn2e}
\usepackage[T1]{fontenc}
\usepackage{url}
\usepackage{amsmath}
\usepackage{txfonts}
\usepackage{mathrsfs}

\renewcommand\mathcal{\mathscr}
\renewcommand\pi\upi
\DeclareMathOperator*{\Res}{Res}

\newcommand\footer[1]{\footnote{\scriptsize #1}}

\newtheorem{theorem}{Theorem}[section]
\newtheorem{corollary}[theorem]{Corollary}

\pagerange{\pageref{start}--\pageref{finish}}
\pubyear{2006}
\journal{{\sc An \& Evans}: {\it The Chang--Refsdal Lens Revisted}}

\begin{document}

\title{The Chang--Refsdal Lens Revisited}
\author[An \& Evans]
{Jin~H.~An$^{1,2}$\thanks{E-mail: jinan@space.mit.edu (JA),
nwe@ast.cam.ac.uk (NWE)} and N.~Wyn~Evans$^2$\footnotemark[1]\\
$^1$MIT Kavli Institute for Astrophysics \& Space Research,
Massachusetts Institute of Technology, 77 Massachusetts Avenue,
Cambridge, MA~02139, USA\\
$^2$Institute of Astronomy, University of Cambridge,
Madingley Road, Cambridge, CB3~0HA, UK}
\date{to appear in {\sl Monthly Notices of Royal Astronomical Society}}

\maketitle
\label{start}
\begin{abstract}
This paper provides a complete theoretical treatment of the point-mass
lens perturbed by constant external shear, often called the
Chang--Refsdal lens. We show that simple invariants exist for the
products of the (complex) positions of the four images, as well as
moment sums of their signed magnifications. The image topographies and
equations of the caustics and critical curves are also studied. We
derive the fully analytic expressions for pre-caustics, which are the
loci of non-critical points that map to the caustics under the lens
mapping. They constitute boundaries of the region in the image domain
that maps onto the interior of the caustics. The areas under the
critical curves, caustics and pre-caustics are all evaluated, which
enables us to calculate the mean magnification of the source within
the caustics. Additionally, the exact analytic expression for the
magnification distribution for the source in the triangular caustics
is derived, as well as a useful approximate expression. Finally, we
find that the Chang--Refsdal lens with additional convergence greater
than unity (the `over-focusing case') can exhibit third-order critical
behaviour, if the `reduced shear' is exactly equal to $\sqrt3/2$, and
that the number of images for $N$ point masses with non-zero constant
shear cannot be greater than $5N-1$.
\end{abstract}

\nokeywords

\section{Introduction}

The so-called Chang--Refsdal lens was put forward to describe the
lensing effects of stars in a background galaxy \citep{CR79,CR84}.
The star acts as a point-mass lens, while the galaxy provides a
background perturbation field, which can be approximated by a tidal
term to lowest order. The Chang--Refsdal lens is therefore the simplest
description of lensing by a tidally perturbed point-mass. As such, it
has found widespread astronomical applications -- e.g. in the
modelling of binary microlensing lightcurves \citep{GL92,GG97l}, in
the analysis of the statistics of high-magnification events caused by
stars in foreground galaxies \citep{Sc87}, and in the study of
microlensing of stars around the black hole in the Galactic Centre
\citep{Al01}, and so on.

The Chang--Refsdal lens is of outstanding physical importance, but it
also has a number of elegant mathematical properties, primarily
because the `deflection function' is a rational function of position
in the lens plane. Although there has already been much work done on
the structure of the caustics and critical curves
\citep{CR84,Su85,Sc92}, many of the properties of the Chang--Refsdal
lens do not appear to be in the literature. The purpose of this paper
is to give a compendium of those results.

The paper is organized as follows. Section \ref{sec:LE} presents the
Chang--Refsdal lens utilizing the complex notation popularized by
\citeauthor{Wi90} (\citeyear{Wi90}; see also \citealt{BK75}).
In section \ref{sec:IP}, the
lens equation is converted into the imaging polynomial, the roots of
which include the image locations. Provided that the source lies
within the caustic -- so that the number of images is four -- then
there exist invariants, such as the products of the image positions or
the sums of the magnifications weighted by powers of the
positions. Section \ref{sec:SSA} considers two cases when the lens
equation is solvable via elementary means, namely when the source lies
on one of the axes of the symmetry. In section \ref{sec:CCC}, the
pre-caustics of the Chang--Refsdal lens are isolated, together with the
critical curves and the caustics. This enables us to calculate in
section \ref{sec:LEM} the mean magnification of the 4-image
configurations, and the exact form for a certain conditional
distribution of magnifications. Finally, sections \ref{sec:CB} and
\ref{sec:NPM} consider two generalizations of the Chang--Refsdal lens,
namely the cases of a convergent background and of $N$ point masses
with shear. We find that, with the convergence greater than unity, the
number of images is either nil or two if the `reduced shear' is
smaller than $\sqrt3/2$, nil, two, or four if it is in between
$\sqrt3/2$ and the unity, and two or four if it is greater than
unity. As for $N$ point masses with shear, we establish that the
maximum number of possible images are bounded by $5N-1$.

\section{The Lens Equation}
\label{sec:LE}

The lowest order effect of the tidal field caused by the external mass
distribution on gravitational lensing is usually described by a
quadratic function that approximates the potential caused by the external
masses. Here, we assume that there is no external mass locally, i.e.,
$\nabla^2\psi_\mathrm{ext}=0$ (zero local convergence; see section
\ref{sec:CB} for the discussion regarding the generalization when
$\nabla^2\psi_\mathrm{ext}=2\kappa\ge0$). Then, the lensing potential
$\psi_\mathrm{ext}$ due to the external mass distribution may be
approximated to be $\psi_\mathrm{ext}(x_1,x_2)=\gamma_\mathrm R
(x_1^2-x_2^2)/2+\gamma_\mathrm Ix_1x_2$, which is actually the most
general form of a quadratic function satisfying Laplace's equation. (The
constant term in the potential has no physical consequence whilst the
linear terms lead to a constant deflection, which can be ignored by
introducing the `offset' between the coordinate origins for the
source and the image position.) If a point-mass lens is present under
the influence of this tidal field, the whole system is described by
the total potential $\psi=(1/2)\ln(x_1^2+x_2^2)+\psi_\mathrm{ext}$,
which is a superposition of the point-mass potential and the external
potential. Here, the coordinate origin in the lens/image plane is
defined by the line of sight towards the lens, and the unit of the
angular measurement is given by the `Einstein ring' radius
corresponding to the mass of the point-mass lens.

Then, since the deflection angle is the gradient of the lensing potential,
we find the lens equation ($\bmath y=\bmath x-\bmath\nabla\psi$):
\begin{equation}
\left\lgroup\begin{array}{c}y_1\\y_2\end{array}\right\rgroup=
\left(1-\frac1{x_1^2+x_2^2}\right)
\left\lgroup\begin{array}{c}x_1\\x_2\end{array}\right\rgroup-
\left\lgroup\begin{array}{cc}\gamma_\mathrm R&\gamma_\mathrm I\\
\gamma_\mathrm I&-\gamma_\mathrm R\end{array}\right\rgroup
\left\lgroup\begin{array}{c}x_1\\x_2\end{array}\right\rgroup,
\label{eq:crlvec}
\end{equation}
or more compactly, utilizing the complex number notation \citep{BK75,Wi90}
\begin{equation}
\zeta=z-\frac1{\bar{z}}-\gamma\bar{z}.
\label{eq:crlz}
\end{equation}
This is usually referred to as the Chang--Refsdal lens, after
\citet{CR79,CR84}. Here, $\bmath y=(y_1,y_2)^\mathrm T$ and $\bmath x=
(x_1,x_2)^\mathrm T$ are the vectors representing the angular
positions of the source (in the absence of the lensing) and of the
image, respectively, whereas $\zeta=y_1+\mathrm iy_2$ and $z=x_1+
\mathrm ix_2$ are their complexified variables. Here, we use the
overbar notation to denote the complex conjugation, e.g. $\bar z=
x_1-\mathrm ix_2$. In the lens equation, the effect of the tidal
field is represented by a symmetric traceless tensor, or equivalently
the complex number $\gamma=\gamma_\mathrm R+\mathrm i\gamma_\mathrm
I$, usually referred to as an external shear. The components of the
(external) shear are related to the (external) potential through
$2\gamma_\mathrm R=(\upartial_{x_1}^2-\upartial_{x_2}^2)
\psi_\mathrm{ext}$ and $\gamma_\mathrm I=\upartial_{x_1}
\upartial_{x_2}\psi_\mathrm{ext}$.

The complex lens equation (\ref{eq:crlz}) can also be directly derived
from the lensing potential by recognizing that the `complex scattering
function' or the `deflection function' is twice the `Wirtinger
derivative' \citep[e.g.][]{SK95} of the lensing potential with
respect to $\bar z$, i.e., the lens equation being $\zeta=z-\alpha$
where $\alpha=2\upartial_{\bar z}\psi$. It is easy to show that this
reduces to equation (\ref{eq:crlz}) with the lensing potential of the
Chang--Refsdal lens given by $\psi=(1/2)\ln\bar z+(1/4)\gamma\bar
z^2+\mbox{C.C.}$, where C.C. indicates the complex conjugate of the
terms in front. In general, the further (Wirtinger) derivatives of
$\alpha$ lead to the convergence $\upartial_z\alpha$ and the total
shear $\upartial_{\bar z}\alpha$ of the system. Of a particular
interest is that, for a null convergent region ($\upartial_z\alpha=
0$), the complex conjugate of the deflection function $\bar\alpha$
becomes analytic \citep{An05}. The introduction of point masses to the
system only adds isolated poles, and thus, the (complex conjugate of
the) deflection function is described by a complex meromorphic (i.e.,
analytic everywhere except at isolated poles) function for any lens
system without a continuous mass distribution. As for the
Chang--Refsdal lens, we find that
\begin{equation}
\zeta=z-\overline{s(z)}\,;\qquad
s(z)=\frac1z+\bar\gamma z,
\label{eq:def}
\end{equation}
where $s(z)$ is not only a meromorphic function but actually a
rational function of $z$ (of degree 2, provided that $\gamma\ne0$).
This implies that the problem of `solving' the lens equation
essentially reduces to finding zeros of a polynomial (see section
\ref{sec:IP}). In addition, we have $\upartial_{\bar z}\alpha=
\overline{\upartial_z\bar\alpha}=\overline{\upartial_zs}=
\overline{s^\prime(z)}=-\bar z^{-2}+\gamma$, that is to say, the total shear
of the Chang--Refsdal lens is simply the sum of the `internal shear'
(i.e., the $-\bar z^{-2}$ term) due to the point mass and the constant
external shear $\gamma$.

We note that it is always possible to choose the direction of the
real axis such that $\gamma$ is a positive real number. In particular,
if $\gamma=|\gamma|\mathrm e^{2\mathrm i\phi_\gamma}$ in a given
coordinate system, the rotation of the coordinate axis by an angle
$\phi_\gamma$ with respect to the origin achieves this. The direction
defined by $\mathrm e^{\mathrm i\phi_\gamma}$ is in fact one of the
eigendirections of the shear tensor, which we shall refer to as the
direction of the shear. The eigenvalue of the shear tensor associated
with the direction of the shear is $|\gamma|$, which is sometimes just
referred to as the shear by itself. Since the shear tensor is real
symmetric and traceless, the second eigendirection (defined by
$\mathrm i\mathrm e^{\mathrm i\phi_\gamma}$) is orthogonal to the
direction of the shear and the associated eigenvalue is given by
`$-|\gamma|$.' In the literature, some authors choose the opposite
sign convention for the tidal effect term in equations (\ref{eq:crlvec})
and (\ref{eq:crlz}), which causes an ambiguity in the definition for
the direction of the shear between the two eigendirections. Our choice
follows the usual weak lensing convention such that the direction of
the image stretching distortion is along the direction of the shear,
but implies that the direction to the mass that causes the tides is
perpendicular to the direction of the tidal shear caused by it.

The Jacobian determinant of equation (\ref{eq:crlvec}), or equivalently,
\begin{equation}
\mathcal J
=\frac{\upartial\zeta}{\upartial z}\frac{\upartial\bar\zeta}{\upartial\bar z}
-\frac{\upartial\zeta}{\upartial\bar z}\frac{\upartial\bar\zeta}{\upartial z}
=1-\left|\frac{\mathrm ds}{\mathrm dz}\right|^2
=1-\left|\frac1{z^2}-\bar\gamma\right|^2
\label{eq:mag}
\end{equation}
defines the area distortion factor under the linearized lens mapping.
Hence, the reciprocal of the Jacobian determinant is (the point-source
limit of) the ratio of the apparent solid angle covering the lensed
image to that of the original source. Since lensing neither creates
nor destroys photons (i.e., the surface brightness is conserved), this
also provides us with the ratio of the apparent flux of the image to
that of the source, in the limit of a point source. When the
individual images corresponding to the given source position are not
resolved, the total magnification (i.e., the flux amplification
factor) is simply given by the sum of individual magnifications over
all images.

The point at which $\mathcal J=0$ is known as the critical point, and
the loci of the critical points are referred to as the critical curves
\citep[e.g.][]{Sc92,Pe01}. In addition, the source position that allows an
image at a critical point is referred to as the caustic point, while
the curves that are the mapping of the critical curves under the lens
equation are the caustics -- that is, a caustic point is the point
on the caustics. The magnification of the image at the
critical point and consequently the source on the caustics is formally
infinite. The caustics are also the boundaries between the source
positions with different number of images. (The converse is not
necessarily true.)

If there exist additional non-critical image positions that also map
onto the caustics under the lens mapping, they are sometimes referred
to as pre-caustic points. The pre-caustics\footnote{Also called the
transition loci by some authors} are the loci of those non-critical
points that map onto the caustics under the lens equation
\citep[e.g.][]{Rh02,Fi02}. The pre-caustics form the boundaries of the
image positions that are the (pre-)images of the interior of the
caustics, whereas the critical curves are the division lines within
them between disjoint domains that the image cannot `cross over'.

\section{The Imaging Polynomial}
\label{sec:IP}

To find the image positions corresponding to a given source position
$\zeta$, equation (\ref{eq:crlz}) needs to be inverted for $z$. The
most straightforward route \citep[c.f.,][for the binary lens equation]
{WM95} involves eliminating $\bar z$ from equation (\ref{eq:crlz}) by means
of its own complex conjugate
\begin{equation}
\bar z=f(z)=\bar\zeta+s(z)=\bar\zeta+\frac1z+\bar\gamma z
=\frac{1+\bar\zeta z+\bar\gamma z^2}z.
\label{eq:econ}
\end{equation}
This results in a rational equation of $z$
\begin{equation}
g(z)=z-\frac1{f(z)}-\gamma f(z)-\zeta
=-\frac{\sum_{k=0}^4a_kz^k}{z(1+\bar\zeta z+\bar\gamma z^2)}=0
\label{eq:impol}
\end{equation}
where
\begin{eqnarray}
a_0&=&\gamma\,,\nonumber\\
a_1&=&2\bar\zeta\gamma+\zeta\,,\nonumber\\
a_2&=&\bar\zeta^2\gamma+2|\gamma|^2+|\zeta|^2\,,\nonumber\\
a_3&=&\bar\gamma\zeta+2\bar\zeta|\gamma|^2-\bar\zeta\,,\nonumber\\
a_4&=&\bar\gamma(|\gamma|^2-1)\,,
\label{eq:impolc}
\end{eqnarray}
that needs to be solved for the image positions. In general, the
polynomials in the numerator and the denominator of $g(z)$ as
expressed in equation (\ref{eq:impol}) are relative prime, and so the
problem reduces to solving a quartic polynomial equation $g_\mathrm
n(z)=\sum_{k=0}^4a_kz^k=0$, which in principle can be solved
algebraically. The result further implies that the number of the
images allowed by equation (\ref{eq:crlz}) for a given source position
cannot be greater than four.

We note however that $g(z_0)=0$ is only the necessary, but not the
sufficient, condition for $z_0$ being one of the image positions. In
other words, not all of the zeros of $g(z)$ are the solution of
equation (\ref{eq:crlz}). For instance, if we replace $\bar z$ with an
unrelated new variable $w$ in equations (\ref{eq:crlz}) and
(\ref{eq:econ}), and solve for $z$ by eliminating $w$, then the
solution is still found from $g(z)=0$ for which $w=f(z)$ is not
necessarily the complex conjugate of $z$.\footnote{However, if
$w_0=f(z_0)\ne\bar z_0$ where $z_0$ is a zero of $g(z)$, then it can
be proven that $\bar w_0$ is another zero of $g(z)$ distinct from the
initial zero $z_0$. That is, any zero of $g(z)$ that does not
correspond to an actual image position always occurs in a pair.}
Therefore, to restrict the zeros of $g(z)$ to actual image positions,
an additional non-algebraic constraint that $\bar z_0=f(z_0)$ where
$g(z_0)=0$ is also required.

We also note that some properties of the Chang--Refsdal lens can be
examined through the analysis of the rational function $g(z)$ and the
quartic polynomial equation $g_\mathrm n(z)=0$. For example,
\begin{equation}
\frac{\mathrm dg}{\mathrm dz}=g^\prime(z)
=1-\left[\gamma-\frac1{f(z)^2}\right]
\left(\bar\gamma-\frac1{z^2}\right)
\label{eq:djac}
\end{equation}
so that, if the solution $z_0$ of $g(z)=0$ corresponds to an actual
image position [i.e., $f(z_0)=\bar z_0$], we find that $g^\prime(z_0)=
\mathcal J(z_0)$. (Here and throughout the paper, the use of the
primed symbols will be reserved for the \emph{total} derivative.)
Since the magnification must be real, this also implies that
$g^\prime(z_0)\in\mathbb R$ is the necessary condition for the zero
$z_0$ of $g(z)$ to correspond to an actual image position. In
addition, if $z_0$ corresponds to a critical point, then $g(z_0)=
g^\prime(z_0)=0$, so that $z_0$ is a degenerate zero of $g(z)$. In
other words, if $\zeta$ is a caustic point, then the imaging equation
$g(z)=0$ allows a degenerate solution that corresponds to a critical
point. Furthermore, $g(z_0)=g^\prime(z_0)=0$ if and only if $g_\mathrm
n(z_0)=g_\mathrm n^\prime(z_0)=0$, and therefore, the discriminant of
the polynomial $g_\mathrm n(z)$ also provides an algebraic equation
$D(\zeta_\mathrm c,\bar\zeta_\mathrm c)=0$ for the caustics points
$\zeta_\mathrm c$ to satisfy.

We further note that, if $z_0$ is the critical point that maps to a
cusp point on the caustics, $g(z_0)=g^\prime(z_0)=g^{\prime\prime}
(z_0)=0$ and $g_\mathrm n(z_0)=g_\mathrm n^\prime(z_0)=g_\mathrm
n^{\prime\prime}(z_0)=0$, and thus that $z_0$ is a doubly-degenerate
solution (i.e., a triple root) of $g(z)=g_\mathrm n(z)=0$. In fact, if
the deflection function is described by an analytic function, we can
establish a connection between multiply-degenerate solutions of a
certain analytic function derived from the deflection function and
higher-order critical behaviour (or `catastrophe') of the lens mapping
at the corresponding point. However, the present analytic function
$g(z)$ can at most have a triple root for any $\zeta\in\mathbb C$ so
that the cusp catastrophe is the critical behaviour of the highest
possible order for the Chang--Refsdal lens, provided that $\gamma\ne0$
and $|\gamma|\ne1$.

Next, for any rational function $r(z)=r_\mathrm n(z)/r_\mathrm d(z)$
of $z$, where $r_\mathrm n(z)$ and $r_\mathrm d(z)$ are polynomials
with no common factor, we find that there exist linear combinations of
linearly-independent polynomials $p_k(z)=[r_\mathrm n(z)]^k[r_\mathrm
d(z)]^{4-k}$ $(k=0,\ldots,4)$ that are divisible by $g_\mathrm n(z)$.
That is, $\sum_{k=0}^4c_kp_k(z)=q(z)g_\mathrm n(z)$ or dividing by
$[r_\mathrm d(z)]^4$,
\begin{equation}
\sum_{k=0}^4c_k
\frac{[r_\mathrm n(z)]^k[r_\mathrm d(z)]^{4-k}}{[r_\mathrm d(z)]^4}=
\sum_{k=0}^4c_k[r(z)]^k=\frac{q(z)g_\mathrm n(z)}{[r_\mathrm d(z)]^4}
\end{equation}
where $q(z)$ is a polynomial of $z$. Thus, for any rational function
$r(z)$, its values $v=r(z_0)$ at the image positions $z_0$ are zeros
of a quartic polynomial $c(v)=\sum_{k=0}^4c_kv^k$, provided that
$r(z)$ does not diverge at the image position [i.e., there is no
common factor between $g_\mathrm n(z)$ and $r_\mathrm d(z)$]. In
particular, if the rational function is given such that $r(z)=R[z,
f(z)]$ where $R[z,w]$ is a rational function of $z$ and $w$ that
satisfies that $R[z,w]\in\mathbb R$ for $w=\bar z$, then only the real
zeros of $c(v)$ correspond to its values at image positions.
Furthermore, if $R[\bar w,\bar z]=\overline{R[z,w]}$, we can show
that, for any non-real zero $v_\mathrm c=r(z_0^\mathrm c)$ of $c(v)$
where $z_0^\mathrm c$ is a zero of $g(z)$ that does not correspond to
any actual image, its complex conjugate $\bar v_\mathrm c$ is also a
zero of $c(v)$. Hence, for this case, $c(v)$ is completely factored
into linear or irreducible quadratic polynomials with real
coefficients, and therefore (after the common scale factor has been
taken out) $c(v)=0$ reduces to a \emph{real-coefficient} quartic
polynomial equation, and only its real solutions are valid, which
correspond to the actual image positions.

For example, if $R[z,w]=zw$, then $R[z,\bar z]=z\bar z=|z|^2\in\mathbb
R$ and $R[\bar w,\bar z]=\bar w\bar z=\overline{zw}=
\overline{R[z,w]}$, and so, with $r(z)=R[z,f(z)]=zf(z)=1+\bar\zeta z+
\bar\gamma z^2$, there exists a real quartic polynomial equation
$\sum_{k=0}^4b_kv^k=0$ whose real solutions are $|z_0|^2$ where $z_0$
are the image positions. By equating the remainder of the polynomial
division of $\sum_{k=0}^4b_k(1+\bar\zeta z+\bar\gamma z^2)^k$ by
$g_\mathrm n(z)$ to be identically nil, we find the real coefficients
(up to a common multiplicative scale factor):
\begin{eqnarray}
b_0&=&1\,,\nonumber\\
b_1&=&-(4+|\zeta|^2)\,,\nonumber\\
b_2&=&6+2(|\zeta|^2-|\gamma|^2)
+\zeta^2\bar\gamma+\bar\zeta^2\gamma\,,\nonumber\\
b_3&=&-4(1-|\gamma|^2)-|\zeta|^2(|\gamma|^2+1)
-(\zeta^2\bar\gamma+\bar\zeta^2\gamma)\,,\nonumber\\
b_4&=&(1-|\gamma|^2)^2\,.
\end{eqnarray}
The result implies that, for the 4-image Chang--Refsdal lens system,
there exists a relation among the image positions that
\begin{equation}
\prod_{k=1}^4|z_k|=\left|\frac1{1-|\gamma|^2}\right|
\end{equation}
where $z_1$, $z_2$, $z_3$, and $z_4$ are the complex locations of four
images. In fact, this relation can be derived from the stricter
relation found from the coefficients of equation (\ref{eq:impolc})
\begin{equation}
\prod_{k=1}^4z_k=\frac\gamma{\bar\gamma(|\gamma|^2-1)}.
\end{equation}

We note that there also exist similar real coefficient quartic
polynomial equations for $2\Re[z_0]$, $2\Im[z_0]$, $\cot(\arg z_0)$,
$\mathcal J(z_0)$ and so on \citep[c.f.,][]{As02,AKK04}. In principle,
they can be found by entirely algebraic operations outlined above with
the choice of the rational function being $z+f(z)$, $\mathrm i
[f(z)-z]$, $\mathrm i[z+f(z)][z-f(z)]^{-1}$, or equation (\ref{eq:djac}),
respectively although the actual algebra involved is rather messy.

\subsection{Position and Magnification Moments}
\label{sec:PMM}

The fact that $g(z_0)=0$ and $g^\prime(z_0)=\mathcal J(z_0)$ for image
positions $z_0$ implies that certain contour integrals involving
$[g(z)]^{-1}$ reduce to the position-magnification moment sum over the
all allowed images \citep{DR01,HE01}, provided that all zeros of
$g(z)$ are the actual image positions. Here, we show that this also
implies that the series of the position-magnification moment sums are
coefficients for particular Taylor-series expansion of $[g(z)]^{-1}$
\citep[see also][]{EH02}.

First, since $\lim_{z\rightarrow\infty}[g(z)]^{-1}=0$ assuming
$|\gamma|\ne1$, we note that $[g(z)]^{-1}$ has the convergent
Taylor-series expansion at $z=\infty$ such that
\begin{equation}
\frac1{g(z)}=\sum_{n=1}^{\infty}\frac{C_n}{z^n}\,;\qquad
C_n=\frac1{n!}\left.\frac{\mathrm d^n}{\mathrm dz^n}
\frac1{(g\circ h)(z)}\right|_{z=0}
\label{eq:pcf}
\end{equation}
where $h(z)\equiv z^{-1}$. Next, let us consider the contour integral
\begin{eqnarray}\lefteqn{
\oint_{\partial S_\infty}\frac{z^n\,\mathrm d z}{g(z)}=
\oint_{\partial S_\infty}\!\mathrm d z\,
z^n\left(\sum_{j=1}^\infty\frac{C_j}{z^j}\right)=
\sum_{j=1}^\infty C_j
\left(\oint_{\partial S_\infty}\!z^{n-j}\mathrm d z\right)
}\nonumber\\*&&
=\left\{\begin{array}{cl}
2\pi\mathrm i C_{n+1}&\mbox{for $n=0,1,\ldots$}\\
0&\mbox{for $n=-1,-2,\ldots$}\end{array}\right.
\label{eq:res}
\end{eqnarray}
where the integration path $\partial S_\infty$ is taken to be a circle
with a radius of infinity. However, the contour integral can also be
evaluated using Cauchy's Residue Theorem. Since $g(0)\ne0$,
non-vanishing residue contributions to the integral are at $z=z_0$
where $z_0$ is a zero of $g(z)$ [i.e., $g(z_0)=0$]
\begin{equation}
\Res_{z=z_0}\left[\frac{z^n}{g(z)}\right]=
\lim_{z\rightarrow z_0}\frac{(z-z_0)z^n}{g(z)}=
\frac{z_0^n}{g^\prime(z_0)}
\end{equation}
for any integer $n$ [also assumed is that none of the zeros of $g(z)$
is degenerate so that $g^\prime(z_0)\ne0$ when $g(z_0)=0$], and at $z=0$
\begin{equation}
\Res_{z=0}\left[\frac1{z^mg(z)}\right]=\frac1{(m-1)!}
\left.\frac{\mathrm d^{m-1}}{\mathrm dz^{m-1}}
\frac1{g(z)}\right|_{z=0}
\label{eq:ncf}
\end{equation}
for a negative integer $n=-m$ ($m=1$, 2, $\ldots$). Then, the residue
theorem, together with equation (\ref{eq:res}) indicates that
\begin{equation}
\sum_k\frac{z_k^n}{g^\prime(z_k)}=\left\{\begin{array}{cl}
C_{n+1}&\mbox{for $n=0,1,\ldots$}\\
-\mbox{Res}_{z=0}\left\{\left[z^{-n}g(z)\right]^{-1}\right\}&
\mbox{for $n=-1,-2,\ldots$}\end{array}\right.,
\label{eq:msm}
\end{equation}
where $z_k\ne0$, $g(z_k)=0$, and $g^\prime(z_k)\ne0$. Here, if all of
the $z_k$'s are the solutions of equation (\ref{eq:crlz}) -- i.e., an
actual image exists at every $z=z_k$ -- then from $g^\prime(z_0)=
\mathcal J(z_0)$, the left-hand side of equation (\ref{eq:msm}) is the same
as the (signed-)magnification-weighted moment sum over the all images.
Furthermore, the formal expressions given in equations (\ref{eq:pcf}) and
(\ref{eq:ncf}) indicate that the sums in equation (\ref{eq:msm}) are also
the coefficients for the Taylor-series expansion of $[g(z)]^{-1}$,
that is to say,
\begin{equation}
\frac1{g(z)}=\sum_{m=1}^\infty
\left.\frac{\mathrm d^m}{\mathrm dw^m}\frac1{g(w^{-1})}\right|_{w=0}
\frac{z^{-m}}{m!}=\sum_{n=0}^\infty\frac1{z^{n+1}}
\left[\sum_k\frac{z_k^n}{g^\prime(z_k)}\right]
\end{equation}
at $z=\infty$, or
\begin{equation}
\frac1{g(z)}=\sum_{n=0}^\infty
\left.\frac{\mathrm d^n}{\mathrm dw^n}\frac1{g(w)}\right|_{w=0}
\frac{z^n}{n!}=
-\sum_{n=0}^\infty z^n\left[\sum_k\frac1{z_k^{n+1}g^\prime(z_k)}\right]
\end{equation}
at $z=0$. We note that this result is reminiscent of Ramanujan's
Master Theorem \citep{Be85}, which relates moment integrals and Taylor
coefficients.

From the series expansion of equation (\ref{eq:impol}) at $z=0$ and
$z=\infty$, we can derived the following invariant position-moment
sums for the Chang--Refsdal lens:
\begin{equation}
\sum_{k=1}^4\frac1{\mathcal J(z_k)}=\frac1{1-|\gamma|^2};
\label{eq:maginv}
\end{equation}
\begin{equation}
\sum_{k=1}^4\frac1{z_k\mathcal J(z_k)}=0\,;\qquad
\sum_{k=1}^4\frac1{z_k^2\mathcal J(z_k)}=\frac1\gamma.
\end{equation}
In addition, we also find that
\begin{equation}
\sum_{k=1}^4\frac{z_k}{\mathcal J(z_k)}=
\frac{\zeta+\gamma\bar\zeta}{(1-|\gamma|^2)^2}\,;\qquad
\sum_{k=1}^4\frac1{z_k^3\mathcal J(z_k)}=
-\frac{\zeta+\gamma\bar\zeta}{\gamma^2},
\end{equation}
which subsequently implies that
\begin{equation}
\sum_{k=1}^4\left[z_k(1-|\gamma|^2)^2+\frac{\gamma^2}{z_k^3}\right]
\frac1{\mathcal J(z_k)}=0.
\end{equation}
Similar results have also been obtained by \citet{Wi00} and \citet{HE01}.

\section{Source on the Symmetry Axis}
\label{sec:SSA}

Although it is in principle possible to solve equation
(\ref{eq:impol}) for the image positions with an algebraic technique
such as Ferrari's method \citep[e.g.][]{AS} for an arbitrary source
position, the actual algebra is rather complicated, if not entirely
uninteresting. In fact, for most cases, applying a simple numerical
algorithm to solve the given quartic polynomial is more advantageous.
However, if the source lies along either of the symmetry axes of the
system, the complete analytic solutions are reasonably simple to
derive and can be illuminating. For the simplicity of the algebra,
throughout this section, we use the lens equation (\ref{eq:crlz})
with the real direction given by the direction of the shear (i.e., the
tide-causing mass lies along the pure imaginary axis) so that $\gamma$
is a positive real number (i.e., $\gamma=\bar\gamma=|\gamma|$).

\subsection{The Real Axis}

\begin{table*}
\caption{
Summary of the image properties for the cases when the source
lies on the real axis (parallel to the direction of the shear). Here,
$A=4\gamma^2-(1+\gamma)\zeta^2$ and $B=\zeta^2+4(1-\gamma)$.}
\label{table:real}
\begin{tabular}{cllc}\hline
Shear & Source Position & Images & Total Magnification $M_\mathrm{tot}$ \\ 
\hline
$0\le\gamma<1$ & $|\zeta|<2\gamma(1+\gamma)^{-1/2}$ & 
4 images [2 on $(+,+)$ \& 2 off $(-,-)$] & 
$\displaystyle\frac{4\gamma-(1+\gamma)\zeta^2}{(1-\gamma^2)A}$
\\
$0\le\gamma<1$ & $|\zeta|>2\gamma(1+\gamma)^{-1/2}$ &
2 images [2 on $(+,-)$] &
$\displaystyle\frac{|\zeta|(\zeta^2-4\gamma+2)}{(1-\gamma)(-A)\sqrt{B}}$
\\
$\gamma>1$ & $|\zeta|<2\sqrt{\gamma-1}$ &
2 images [2 off $(-,-)$] &
$\displaystyle\frac{2\gamma}{(\gamma+1)A}$
\\
$\gamma>1$ & $2\sqrt{\gamma-1}<|\zeta|<2\gamma(\gamma+1)^{-1/2}$ &
4 images [2 on $(+,-)$ \& 2 off $(-,-)$] &
$\displaystyle\frac{|\zeta|(4\gamma-2-\zeta^2)}{(\gamma-1)A\sqrt{B}}
+\frac{2\gamma}{(\gamma+1)A}$
\\
$\gamma>1$ & $|\zeta|>2\gamma(\gamma+1)^{-1/2}$ &
2 images [2 on $(-,-)$] &
$\displaystyle\frac{\zeta^2-2\gamma}{(\gamma-1)(-A)}$
\\\hline
\end{tabular}
\end{table*}

First, let us think of the case when the source is along the real axis
(i.e., $\bar\zeta=\zeta$). From the symmetry of the problem, we expect
that there will be images along the real axis, too. Since any solution
along the real axis satisfies $z=\bar z$ and equation (\ref{eq:crlz}),
we can derive the quadratic equation for $z$
\begin{equation}
g_\mathrm R(z)=(1-\gamma)z^2-\zeta z-1=0.
\label{eq:quad}
\end{equation}
It is a straightforward exercise to show that $g_\mathrm R(z)$ is a
factor of $g_\mathrm n(z)$ if $\zeta=\bar\zeta$ and $\gamma=
\bar\gamma$. In particular, we find that
\begin{equation}
g_\mathrm n(z)=-g_\mathrm R(z)
\left[\gamma(1+\gamma)z^2+(1+\gamma)\zeta z+\gamma\right]
\label{eq:quadfac}
\end{equation}
and therefore, the solutions of the quartic equation $g_\mathrm n(z)=
0$ are 
\begin{eqnarray}\lefteqn{
z_\mathrm{on}=
\frac{\zeta\pm\left[\zeta^2+4(1-\gamma)\right]^{1/2}}{2(1-\gamma)},
}\nonumber\\\label{eq:rs}\lefteqn{
z_\mathrm{off}=
-\frac\zeta{2\gamma}\pm\frac{\mathrm i}{2\gamma}
\left(\frac{4\gamma^2}{1+\gamma}-\zeta^2\right)^{1/2}.
}\end{eqnarray}
Here, two on-axis solutions $z_\mathrm{on}$ are image locations if
$B\equiv\zeta^2+4(1-\gamma)\ge0$. Since $\zeta\in\mathbb R$ and
$\gamma\ge0$, this happens (i) if $0\le\gamma<1$ or (ii) if $\gamma>1$
and $|\zeta|\ge2\sqrt{\gamma-1}$. On the other hand, the two off-axis
solutions $z_\mathrm{off}$ become physical image positions if $|\zeta|
\le2\gamma(1+\gamma)^{-1/2}$. The corresponding magnifications of the
images can be found from equation (\ref{eq:mag}). We remark that the
algebra may be eased by reducing the orders of the power of $z$
through the use of quadratic equations (\ref{eq:quad}) and
(\ref{eq:quadfac}) for image positions and by deriving additional
quadratic equations for magnification via a similar route using the
resultant polynomials, as in section \ref{sec:IP}. Furthermore, since
equation (\ref{eq:mag}) is invariant under complex conjugation,
and the two off-axis solutions are mutual complex conjugates, they
must have identical magnifications. We find that the magnification for
each off-axis image is
\begin{equation}
M_\mathrm{off}=-\frac\gamma{(1+\gamma)A}
\label{eq:rmo}
\end{equation}
where $A\equiv4\gamma^2-(1+\gamma)\zeta^2$. If $|\zeta|<2\gamma
(1+\gamma)^{-1/2}$, this is negative so that both off-axis images are
of negative parity. If $|\zeta|=2\gamma(1+\gamma)^{-1/2}$, the
magnification diverges, implying that $\zeta=\pm2\gamma
(1+\gamma)^{-1/2}$ are the caustic points along the real axis. The
resulting expressions for the magnifications of the on-axis images are
rather complicated, but their sum, product, and difference, which can
be derived using resultant quadratic polynomials, are simpler to write
down;
\begin{eqnarray}\lefteqn{
M_1+M_2=-\frac{\zeta^2-2\gamma}{(1-\gamma)A}\,;\qquad
M_1M_2=\frac1{(1-\gamma)AB}\,;}\nonumber\\\label{eq:rmt}\lefteqn{
|M_1-M_2|=\frac{|\zeta||\zeta^2-4\gamma+2|}{|1-\gamma||A|\sqrt{B}},}
\end{eqnarray}
where $B\equiv\zeta^2+4(1-\gamma)$ as defined following equation
(\ref{eq:rs}). With $B>0$, the two on-axis images
are of the same parity if $(1-\gamma)A>0$, whereas they are of opposite
parity if $(1-\gamma)A<0$. In particular, if $0\le\gamma<1$ and
$|\zeta|<2\gamma(1+\gamma)^{-1/2}$, both are positive parity images,
whilst both are negative parity images if $\gamma>1$ and $|\zeta|>2
\gamma(1+\gamma)^{-1/2}$. Again, if $\zeta=\pm2\sqrt{\gamma-1}$ with
$\gamma>1$ or if $\zeta=\pm2\gamma(1+\gamma)^{-1/2}$, the
magnification of either of the on-axis images diverges, so that the
source location corresponds to the caustic point. The total
magnification can be found from the sum of the absolute values of the
individual magnifications corresponding to the actual image positions.
The results for all these case are summarized in Table
\ref{table:real}. We find that two negative off-axis images and one
positive on-axis image merge into one negative on-axis image at
$\zeta=\pm2\gamma(\gamma+1)^{-1/2}$, whereas two opposite parity
on-axis images appear or disappear simply by merging with each other
at $\zeta=\pm2\sqrt{\gamma-1}$. Note that this reflects that the formers
are actually cusp points while the latters are simple first-order
caustic points.

\subsection{The Imaginary Axis}

\begin{table*}
\caption{
Summary of the image properties for the cases when the source
lies on the imaginary axis (perpendicular to the direction of the
shear or parallel to the direction of the tide-causing mass). Here,
$C=4\gamma^2-(1-\gamma)|\zeta|^2=4\gamma^2+(\gamma-1)|\zeta|^2$ and
$D=|\zeta|^2+4(1+\gamma)$.}
\label{table:imag}
\begin{tabular}{cllc}\hline
Shear & Source Position & Images & Total Magnification $M_\mathrm{tot}$ \\ 
\hline
$0\le\gamma<1$ & $|\zeta|<2\gamma(1-\gamma)^{-1/2}$ &
4 images [2 on $(-,-)$ \& 2 off $(+,+)$] &
$\displaystyle\frac{4\gamma+(1-\gamma)|\zeta|^2}{(1-\gamma^2)C}$
\\
$0\le\gamma<1$ & $|\zeta|>2\gamma(1-\gamma)^{-1/2}$ &
2 images [2 on $(+,-)$]&
$\displaystyle\frac{|\zeta|(|\zeta|^2+4\gamma+2)}{(1+\gamma)(-C)\sqrt{D}}$
\\
$\gamma>1$ & any & 2 images [2 on $(-,-)$]&
$\displaystyle\frac{|\zeta|^2+2\gamma}{(\gamma+1)C}$
\\\hline
\end{tabular}
\end{table*}

The lensing of the source along the imaginary axis (i.e., $\bar\zeta=
-\zeta$) can be analyzed in a similar manner. First, by considering
the solutions on the imaginary axis ($\bar z=-z$),
we can easily factor $g_\mathrm n(z)$ into two quadratics if
$\bar\zeta=-\zeta$ and $\bar\gamma=\gamma$, and subsequently find the
zeros of $g_\mathrm n(z)$:
\begin{eqnarray}\lefteqn{
z_\mathrm{on}=\mathrm i\
\frac{\Im[\zeta]\pm\left[|\zeta|^2+4(1+\gamma)\right]^{1/2}}{2(1+\gamma)}
}\nonumber\\\label{eq:is}\lefteqn{
z_\mathrm{off}=
\frac{\mathrm i\,\Im[\zeta]}{2\gamma}\pm\frac1{2\gamma}
\left(\frac{4\gamma^2}{1-\gamma}-|\zeta|^2\right)^{1/2}.
}\end{eqnarray}
Here, $\zeta=\mathrm i\Im[\zeta]$ with $\Im[\zeta]\in\mathbb R$ and
$\zeta^2=-|\zeta|^2\le0$. With $\gamma\ge0$, the two on-axis zeros
$z_\mathrm{on}$ are always image positions. On the other hand, the two
off-axis zeros $z_\mathrm{off}$ become image positions if $0\le\gamma<
1$ and $|\zeta|<2\gamma(1-\gamma)^{-1/2}$. The two off-axis solutions
are related to each other by $z\leftrightarrow-\bar z$, which is also
a transformation that leaves equation (\ref{eq:mag}) invariant, and so
they have identical magnifications too, namely
\begin{equation}
M_\mathrm{off}=\frac\gamma{(1-\gamma)C},
\label{eq:imo}
\end{equation}
where $C\equiv4\gamma^2-(1-\gamma)|\zeta|^2$. We find $M_\mathrm{off}$
to be positive if $|\zeta|<2\gamma(1-\gamma)^{-1/2}$ with $0\le\gamma
<1$. The sum, product, and difference of the two on-axis images are
similarly found to be
\begin{eqnarray}\lefteqn{
M_1+M_2=-\frac{|\zeta|^2+2\gamma}{(1+\gamma)C}\,;\qquad
M_1M_2=\frac1{(1+\gamma)CD}\,;}\nonumber\\\label{eq:imt}\lefteqn{
|M_1-M_2|=\frac{|\zeta|(|\zeta|^2+4\gamma+2)}{(1+\gamma)|C|\sqrt{D}},}
\end{eqnarray}
where $D\equiv|\zeta|^2+4(1+\gamma)$. This indicates that the two
on-axis images are of opposite parity if $C<0$ and that they are both
negative parity images if $C>0$. Along the imaginary axis, the caustic
points occur at $\zeta=\pm2\mathrm i\gamma(1-\gamma)^{-1/2}$ with $0
\le\gamma<1$, for which two positive off-axis images merges with one
negative on-axis image into one positive on-axis image. We provide a
summary of these results including the total magnification in Table
\ref{table:imag}.

We note that the case of the source on the imaginary axis with a
positive real shear is actually mathematically identical to the case
of the source along the real axis with a \emph{negative} real shear,
modulo a 90\degr\ rotation of the system. In fact, equations
(\ref{eq:is}-\ref{eq:imt}) can be obtained from equations
(\ref{eq:rs}-\ref{eq:rmt}) through the transformation
$\gamma\rightarrow-\gamma$ (and $z\rightarrow\mathrm
iz$). Nonetheless, we chose to present the case when the source lies
on the symmetry axis that is perpendicular to the shear separately
from that on the axis that is parallel, in order to clarify the
physical difference between the two cases.

\subsection{The Magnification along the Symmetry Axis}

\begin{figure}
\includegraphics[width=\hsize]{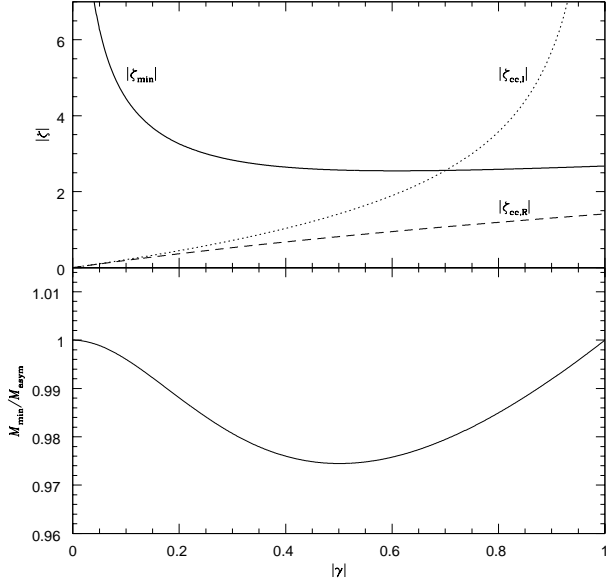}
\caption{\label{fig:min}
Top: The location of the global minimum
magnification for the source on the real axis for $0\le\gamma<1$
(solid line). Also plotted are the locations of the caustic points (in
fact, cusp points) along the real axis (dashed line) and along the
imaginary axis (dotted line). Bottom: The ratio of the global minimum
magnification $M_{\min}$ to the asymptotic magnification
$M_\mathrm{asym}=(1-\gamma^2)^{-1}$ at infinity.}
\end{figure}

\begin{figure}
\includegraphics[width=\hsize]{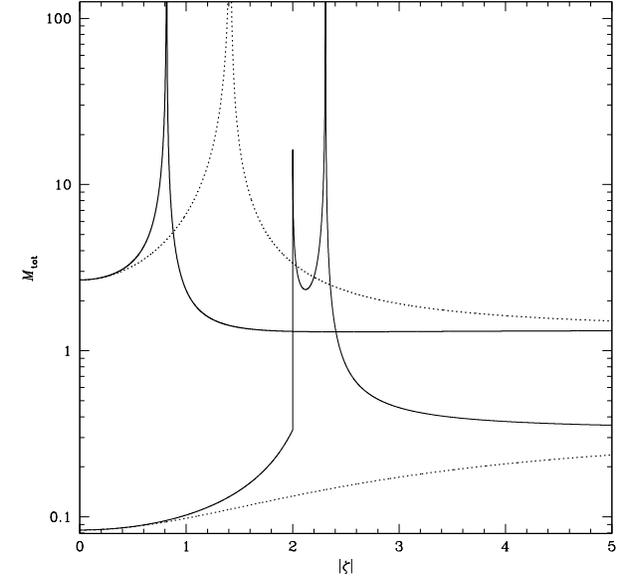}
\caption{\label{fig:mag}
Magnification along the symmetry axis. Thick
lines are for $\gamma=1/2$ while thin lines are for $\gamma=2$. The
solid lines are along the real axis (parallel to the direction of the
shear), and the dotted lines are along the imaginary axis
(perpendicular to the direction of the shear).}
\end{figure}

For $0\le\gamma<1$, the total magnification of the source at the
centre is $M_\mathrm{tot}=[\gamma(1-\gamma^2)]^{-1}$. As the source
moves along the symmetry axis, the magnification increases until the
source crosses the caustics where the magnification diverges. The
caustic points on the real axis are located at $\zeta=\pm2\gamma
(1+\gamma)^{-1/2}$, while the ones on the imaginary axis are at
$\zeta=\pm2\mathrm i\gamma(1-\gamma)^{-1/2}$. So, we deduce that the
caustic is elongated along the imaginary axis (i.e., perpendicular to
the shear direction or parallel to the direction of the tide-causing
mass). The magnification initially decreases from infinity as the
source moves past the caustic points along the symmetry axis. Along
the imaginary axis, it decreases monotonically and tends to
$(1-\gamma^2)^{-1}$ at infinity. Along the real axis, however, the
global minimum magnification, which is slightly smaller than the
asymptotic magnification, occurs at some finite distance from the
centre, and the magnification increases back to the asymptotic value
$(1-\gamma^2)^{-1}$ at infinity.
The location and the value of the global minimum magnification (as a
function of $\gamma$) can in principle be derived from the expression
for the total magnification given in Table \ref{table:real};
\begin{eqnarray}\lefteqn{
\zeta_{\min}^2=4\gamma-1+\frac{1+E}\gamma\,;\qquad
E=\sqrt{1-2\gamma+9\gamma^2}
}\nonumber\\\lefteqn{
M_{\min}=\frac{1+\gamma+E}{(1-\gamma)[1+3\gamma^2+(1+\gamma)E]}
\sqrt{\frac{1-\gamma+4\gamma^2+E}{1+3\gamma+E}}.}
\end{eqnarray}
Figure \ref{fig:min} shows $|\zeta_{\min}|$ and $M_{\min}/
M_\mathrm{asym}$ as functions of $\gamma$ where $M_\mathrm{asym}=
(1-\gamma^2)^{-1}$ is the asymptotic magnification. We find that
$M_{\min}>1$ and that $M_{\min}\approx M_\mathrm{asym}$. In fact,
for $0<\gamma<1$, we have $1>M_{\min}/M_\mathrm{asym}\ge9\sqrt3/16
\approx0.97428$ where the minimum of $M_{\min}/M_\mathrm{asym}$
occurs at $\gamma=1/2$.
In Fig.~\ref{fig:mag}, we plot the behaviour of the total
magnification, as the source moves along the symmetry axis for
$\gamma=1/2$.

For $\gamma>1$, the total magnification for the source at the centre
is given by $[2\gamma(\gamma+1)]^{-1}$ (which is actually the global
minimum magnification), so that the lensed images are in fact
demagnified compared to the source. The magnification increases as the
source moves along the symmetry axis. Along the imaginary axis, the
magnification monotonically increases throughout and tends to the
asymptotic value $(\gamma^2-1)^{-1}$ at infinity. Along the real
axis, there is a discontinuous jump in the magnification to infinity
when the source crosses the caustic at $\zeta=\pm2\sqrt{\gamma-1}$. As
the source moves across the caustic, the magnification drops from
infinity to a local minimum value and then diverges back to infinity
as the source approaches the other caustic points
$\zeta=\pm2\gamma(\gamma+1)^{-1/2}$ along the real axis. Once the
source moves out of the caustic, the magnification monotonically
decreases from infinity and tends to the same asymptotic value
$(\gamma^2-1)^{-1}$. Here, the local minimum value of the
magnification and the corresponding source position within the caustic
are actually easy to find. First, we note that there are four images
-- one positive and three negative parity -- for the case considered
here. However, from the signed magnification sum invariant given in
equation (\ref{eq:maginv}), we find that the total magnification is
completely specified by the magnification $M_+$ of the positive parity
image alone, that is, $M_\mathrm{tot}=2M_+-(1-\gamma^2)^{-1}$.
Finally, from equation (\ref{eq:mag}), the minimum possible magnification
of any positive parity image is unity when $z^2=\bar\gamma^{-1}$. With
$\bar\gamma=\gamma>1$, we find that $z=\pm\gamma^{-1/2}$ maps to
$\zeta=\mp\gamma^{-1/2}(2\gamma-1)$ so that we in fact have
$2\sqrt{\gamma-1}<|\zeta|<2\gamma (\gamma+1)^{-1/2}$ and the
corresponding local minimum magnification is given by $M_\mathrm{tot}=
(2\gamma^2-1)/(\gamma^2-1)$. In Fig.~\ref{fig:mag}, we also plot the
total magnification for the source on the symmetry axis for
$\gamma=2$.

\section{Critical Curves and Caustics}
\label{sec:CCC}

From equation (\ref{eq:mag}), the parametric representation of the critical
curve can be found \citep{Wi90} by solving
\begin{equation}
\frac1{z_\mathrm c^2}-\bar\gamma=\mathrm e^{-2\mathrm i\phi}
\qquad\rightarrow\qquad
z_\mathrm c^2=\frac1{\bar\gamma+\mathrm e^{-2\mathrm i\phi}}
=\frac{\mathrm e^{2\mathrm i\phi}}{1+\bar\gamma\mathrm e^{2\mathrm i\phi}}.
\label{eq:crisq}
\end{equation}
Here, the closed form solutions for the critical curve involve the
square-root, but because of the presence of the branch-cut in the
mapping of $z^{1/2}$, we need to be careful. Once the branch-cut
of the mapping $z^{1/2}$ is assumed to lie along the negative real
axis, the resulting expressions for the critical curve and the caustic
should differ depending on the magnitude of $|\gamma|$ in order to
avoid crossing the branch-cut (i.e., there is no discontinuous jump in
the resulting parametric expression). Hence, we find the parametric
representations of the critical curves
\begin{equation}
z_\mathrm c(\phi)=
\frac{\mathrm e^{\mathrm i\phi}}
{\left(1+\bar\gamma\mathrm e^{2\mathrm i\phi}\right)^{1/2}}=
\frac{\mathrm e^{\mathrm i\tilde\phi}}
{\left(1+|\gamma|\mathrm e^{2\mathrm i\tilde\phi}\right)^{1/2}}\
\mathrm e^{\mathrm i\phi_\gamma}
\label{eq:cri1}
\end{equation}
for $0\le|\gamma|<1$, and
\begin{equation}
z_\mathrm c(\phi)=
\pm\frac1{\bar\gamma^{1/2}
\left(1+\bar\gamma^{-1}\mathrm e^{-2\mathrm i\phi}\right)^{1/2}}=
\pm\frac1{\left(|\gamma|+\mathrm e^{-2\mathrm i\tilde\phi}\right)^{1/2}}\
\mathrm e^{\mathrm i\phi_\gamma}
\label{eq:cri2}
\end{equation}
for $|\gamma|>1$. Here, $\gamma=|\gamma|\mathrm e^{2\mathrm i
\phi_\gamma}$ and $\tilde\phi=\phi-\phi_\gamma$. The parametric
representation of the caustics can be found by mapping the critical
curve though the lens equation (\ref{eq:crlz});
\begin{equation}
\zeta_\mathrm c(\phi)=z_\mathrm c(\phi)
-\frac1{\bar z_\mathrm c(\phi)}-\gamma\bar z_\mathrm c(\phi).
\end{equation}

However, before we actually derive expressions, let us examine first
\begin{eqnarray}\lefteqn{
\zeta_\mathrm c^\prime(\phi)=
z_\mathrm c^\prime(\phi)\upartial_z\zeta
+\bar z_\mathrm c^\prime(\phi)\upartial_{\bar z}\zeta
=z_\mathrm c^\prime(\phi)+\mathrm e^{2\mathrm i\phi}
\bar z_\mathrm c^\prime(\phi)
}\nonumber\\*&&=\mathrm e^{\mathrm i\phi}
\left[\mathrm e^{-\mathrm i\phi}z_\mathrm c^\prime(\phi)
+\mathrm e^{\mathrm i\phi}\bar z_\mathrm c^\prime(\phi)\right]
=2\mathrm e^{\mathrm i\phi}
\Re[\mathrm e^{-\mathrm i\phi}z_\mathrm c^\prime(\phi)].
\end{eqnarray}
In other words, we find that $\zeta_\mathrm c^\prime(\phi)$ is
parallel to $\mathrm e^{\mathrm i\phi}$, that is, the parameter $\phi$
is defined such that it is the argument angle of the tangent to the caustics at
the caustic point. Furthermore, we also find that the caustics form a
cusp point [$\zeta_\mathrm c^\prime(\phi)=0$] if $\mathrm e^{-\mathrm
i\phi} z_\mathrm c^\prime(\phi)$ is pure imaginary number, that is
$z_\mathrm c^\prime(\phi)=\mathrm i r\mathrm e^{\mathrm i\phi}$ with
$r\in\mathbb R$. This indicates that $z_\mathrm c^\prime(\phi)$ is
orthogonal to $\mathrm e^{\mathrm i\phi}$ at the critical point
corresponding to a cusp point.
In fact, the general condition for the cusp along the caustics is
that the tangent to the caustic is the normal direction to the
critical curve at the corresponding caustic point and the critical
point.

Now, the actual parametric expression of the caustics of the
Chang--Refsdal lens are given by
\begin{eqnarray}\lefteqn{
\zeta_\mathrm c(\phi)=\mathrm e^{\mathrm i\phi}\left[
\frac1{\left(1+\bar\gamma\mathrm e^{2\mathrm i\phi}\right)^{1/2}}
-\frac{1+2\gamma\mathrm e^{-2\mathrm i\phi}}
{\left(1+\gamma\mathrm e^{-2\mathrm i\phi}\right)^{1/2}}\right]
}\nonumber\\*&&=
\mathrm e^{\mathrm i\tilde\phi}\left[
\frac1{\left(1+|\gamma|\mathrm e^{2\mathrm i\tilde\phi}\right)^{1/2}}
-\frac{1+2|\gamma|\mathrm e^{-2\mathrm i\tilde\phi}}
{\left(1+|\gamma|\mathrm e^{-2\mathrm i\tilde\phi}\right)^{1/2}}\right]\
\mathrm e^{\mathrm i\phi_\gamma}
\end{eqnarray}
for $0\le|\gamma|<1$, and
\begin{eqnarray}\lefteqn{
\zeta_\mathrm c(\phi)=\pm\frac1{\bar\gamma^{1/2}}
\left[\frac1{\left(1+\bar\gamma^{-1}\mathrm e^{-2\mathrm i\phi}\right)^{1/2}}
-|\gamma|\frac{2+\gamma^{-1}\mathrm e^{2\mathrm i\phi}}
{\left(1+\gamma^{-1}\mathrm e^{2\mathrm i\phi}\right)^{1/2}}\right]
}\nonumber\\*&&=
\pm\left[\frac1{\left(|\gamma|+\mathrm e^{-2\mathrm i\tilde\phi}\right)^{1/2}}
-\frac{2|\gamma|+\mathrm e^{2\mathrm i\tilde\phi}}
{\left(|\gamma|+\mathrm e^{2\mathrm i\tilde\phi}\right)^{1/2}}\right]\
\mathrm e^{\mathrm i\phi_\gamma}
\end{eqnarray}
for $|\gamma|>1$. Here, we note that the critical curves and the
caustics are single simply-connected curves with $\phi\in[0,2\pi)$ for
$0\le|\gamma|<1$, while they are two separate mirror-symmetric closed
curves with $\phi\in[0,\pi)$ for $\gamma>1$.

The cusp points along the caustics can be found using
\begin{equation}
\zeta_\mathrm c^\prime(\phi)=
2\mathrm e^{\mathrm i\phi}
\Im\left[\frac1{\left(1+\gamma\mathrm e^{-2\mathrm i\phi}\right)^{3/2}}\right]
\end{equation}
for $0\le|\gamma|<1$, and
\begin{equation}
\zeta_\mathrm c^\prime(\phi)=
\pm2\mathrm e^{\mathrm i\phi}
\Im\left[\frac{\mathrm e^{3\mathrm i\phi}}
{\left(\gamma+\mathrm e^{2\mathrm i\phi}\right)^{3/2}}\right]
\end{equation}
for $|\gamma|>1$. First, for $0\le|\gamma|<1$, we find that the
condition $\zeta_\mathrm c^\prime(\phi)=0$ is equivalent to $(1+
|\gamma|\mathrm e^{2\mathrm i\tilde\phi})^{3/2}\in\mathbb R$ so that
we have $2\mathrm i\tilde\phi=n\pi$ for $\tilde\phi$ corresponding to
the cusp points. Here, $n$ is an integer. Hence, the caustics have
four cusps at
\begin{equation}
\zeta_{\mathrm c\mathrm c}=\pm\mathrm e^{\mathrm i\phi_\gamma}
\frac{2|\gamma|}{\left(1+|\gamma|\right)^{1/2}}
\,;\qquad
\zeta_{\mathrm c\mathrm c}=\pm\mathrm i\mathrm e^{\mathrm i\phi_\gamma}
\frac{2|\gamma|}{\left(1-|\gamma|\right)^{1/2}},
\end{equation}
which is consistent with the result in the previous section. Next, for
$|\gamma|>1$, the condition that $\zeta_\mathrm c^\prime(\phi)=0$
reduces to $\left(|\gamma|\mathrm e^{2\mathrm i\tilde\phi}+1\right)^3$
being a \emph{positive} real number, or equivalently $\arg(1+|\gamma|
\mathrm e^{2\mathrm i\tilde\phi})=(2n\pi)/3$ where $n$ is again an
integer. From simple geometric considerations, this reduces to
$2\tilde\phi=2n\pi$, $2n\pi+(2\pi)/3+\xi$, or $2n\pi+(4\pi)/3-\xi$
where $\xi=\arcsin[\sqrt3/(2|\gamma|)]$.
%
%
Therefore, we find that there are three cusps along each caustic
(hence, in total six, $2\times3=6$) at
\begin{eqnarray}\lefteqn{
\zeta_{\mathrm c\mathrm c}=
\pm\mathrm e^{\mathrm i\phi_\gamma}\frac{\sqrt2\mathrm e^{\mathrm i\xi/2}}
{\left(\sqrt{4|\gamma|^2-3}-1\right)^{1/2}}
\left(2-\sqrt{4|\gamma|^2-3}+\mathrm e^{2\mathrm i\pi/3}\right)
}\nonumber\\\lefteqn{
\zeta_{\mathrm c\mathrm c}=
\pm\mathrm e^{\mathrm i\phi_\gamma}\frac{\sqrt2\mathrm e^{-\mathrm i\xi/2}}
{\left(\sqrt{4|\gamma|^2-3}-1\right)^{1/2}}
\left(2-\sqrt{4|\gamma|^2-3}+\mathrm e^{-2\mathrm i\pi/3}\right)
}\nonumber\\\lefteqn{
\zeta_{\mathrm c\mathrm c}=
\mp\mathrm e^{\mathrm i\phi_\gamma}
\frac{2|\gamma|^{1/2}}{\left(1+|\gamma|^{-1}\right)^{1/2}}.
}\end{eqnarray}

\subsection{The Areas under the Critical Curves and the Caustics}

Next, since $\mathrm e^{2\mathrm i\phi}$ is periodic for $\phi$ (with
period $\pi$), one can conclude that $z_\mathrm c(\phi)$ [and
consequently $\zeta_\mathrm c(\phi)$] is also periodic although the
period is not necessarily $\pi$. If there exists a convergent
Fourier-series expansion
\begin{equation}
z_\mathrm c(\phi)=\sum_{k=-\infty}^\infty
c_k\mathrm e^{\mathrm ik\omega\phi}\,;\qquad
\zeta_\mathrm c(\phi)=\sum_{k=-\infty}^\infty
d_k\mathrm e^{\mathrm ik\omega\phi}
\end{equation}
where $\omega$ is the angular frequency of $z_\mathrm c(\phi)$ [i.e.,
the $\phi$-period of $z_\mathrm c(\phi)$ is $2\pi/\omega$], then from
\begin{eqnarray}\lefteqn{
\zeta_\mathrm c^\prime(\phi)=z_\mathrm c^\prime(\phi)
+\mathrm e^{2\mathrm i\phi}\bar z_\mathrm c^\prime(\phi)=
\sum_k\mathrm ik\omega c_k\mathrm e^{\mathrm ik\omega\phi}
-\sum_k\mathrm ik\omega\bar c_k\mathrm e^{\mathrm i(2-k\omega)\phi}
}\nonumber\\*&&=\sum_k\mathrm i\omega
\left[kc_k+\left(k-\frac2\omega\right)\bar c_{2/\omega-k}\right]
\mathrm e^{\mathrm ik\omega\phi},
\end{eqnarray}
we find that $kd_k=kc_k+(k-2/\omega)\bar c_{2/\omega-k}$, assuming
$2/\omega$ is an integer [i.e, the $\phi$-period of $z_\mathrm c
(\phi)$ is an integer multiple of $\pi$]. Here, the relations for
$k=0$ and $k=2/\omega$ together indicate that $c_{2/\omega}=
d_{2/\omega}=0$, whereas $c_0$ and $d_0$ are unspecified. Then the
caustics may be written in terms of $c_k$'s as
\begin{eqnarray}\lefteqn{
\zeta_\mathrm c(\phi)-d_0=\sum_{k\ne0}
\left[c_k+\frac{k-2/\omega}k\bar c_{2/\omega-k}\right]
\mathrm e^{\mathrm ik\omega\phi}
}\nonumber\\*&&=\sum_{k\ne2/\omega}
\frac{(k-2/\omega)c_k\mathrm e^{\mathrm ik\omega\phi}
+k\bar c_k\mathrm e^{-\mathrm i(k\omega-2)\phi}}{k-2/\omega}-c_0.
\end{eqnarray}

One application for which the Fourier-series expansions may be useful
is to find the area within the critical curve and the caustics.
We note that the `area 2-form' in $\mathbb C$ may be written as
$\mathrm d^2\!\bmath x=\mathrm d(\bar z\mathrm dz)/
(2\mathrm i)$.\footnote{$\mathrm d(\bar z\mathrm dz)=
\mathrm d\bar z\wedge\mathrm dz=
(\mathrm dx_1-\mathrm i\mathrm dx_2)\wedge
(\mathrm dx_1+\mathrm i\mathrm dx_2)=
2\mathrm i\ \mathrm dx_1\wedge\mathrm dx_2$}
Then, the (signed) area within a simply-connected closed curve
$\partial C=\{z(p)|z(p):[0,P)\rightarrow\mathbb C, p\in[0,P)\subset
\mathbb R\}$ where $p$ is a parameter and $P$ is the period can be
found to be, from the fundamental theorem of multivariate calculus
-- the `generalized Stoke's theorem' \citep[c.f.,][]{GG97,Do98}
\begin{equation}
S=
\int_C\!\mathrm d^2\!\bmath x=
\frac1{2\mathrm i}\int_C\!\mathrm d(\bar z\mathrm dz)=
\frac1{2\mathrm i}\oint_{\partial C}\!\bar z\,\mathrm dz=
\frac1{2\mathrm i}\int_0^P\!\bar z(p)\,z^\prime(p)\,\mathrm dp.
\label{eq:area}
\end{equation}
The area within the critical curve can be expressed in terms of its
Fourier coefficients, from the generalized Parseval's theorem, as
\begin{equation}
S[z(\phi)]=\frac1{2\mathrm i}\int_0^{2\pi/\omega}
\bar z_\mathrm c(\phi)\,z_\mathrm c^\prime(\phi)\,\mathrm d\phi
=\pi\sum_kk|c_k|^2.
\end{equation}
%
Provided (i) that the period of $\zeta_\mathrm c(\phi)$ is also
$2\pi/\omega$ and (ii) that the caustic $\zeta_\mathrm c(\phi)$ does
not exhibit self-intersection, the area within the caustic can also be
written in similar manner
\begin{eqnarray}\lefteqn{
\frac{S[\zeta_\mathrm c(\phi)]}\pi=\sum_kk|d_k|^2=\sum_{k\ne0}\frac1k
\left|kc_k+\left(k-\frac2\omega\right)\bar c_{2/\omega-k}\right|^2
}\nonumber\\*&&=\sum_{k\ne2/\omega}\frac{2k}{2-k\omega}|c_k|^2
-\frac2\omega\sum_k\Re[c_kc_{2/\omega-k}].
\end{eqnarray}
We note that this is typically negative as the orientation chosen by
the parameter $\phi$ is usually clockwise.

For $0\le|\gamma|<1$, from equation (\ref{eq:cri1}), we find the
Fourier-series expansions of the critical curves and the caustics to be
\begin{equation}
z_\mathrm c\mathrm e^{-\mathrm i\phi_\gamma}=\sum_{k=0}^\infty
\frac{(-1)^k}{k!}\left(\frac12\right)_k
|\gamma|^k\mathrm e^{(2k+1)\mathrm i\tilde\phi}
\end{equation}
\begin{equation}
\zeta_\mathrm c\mathrm e^{-\mathrm i\phi_\gamma}=
\sum_{k=1}^\infty
\frac{C_k}2\left[(2k-1)\mathrm e^{(2k+1)\mathrm i\tilde\phi}
+(2k+1)\mathrm e^{-(2k-1)\mathrm i\tilde\phi}\right]|\gamma|^k
\end{equation}
where $(a)_k$ is Pochhammer symbol such that
\begin{equation}
\left(a\right)_k\equiv a(a+1)\cdots(a+k-1)
=\frac{\Gamma(a+k)}{\Gamma(a)}
\end{equation}
and
\begin{equation}
C_k=(-1)^k\frac{\Gamma(k-1/2)}{k!\Gamma(1/2)}.
\label{eq:ck}
\end{equation}
We note that both series expansions are also Taylor-series expansions
with respect to $\gamma$ at $\gamma=0$. If we truncate the expansion
of the caustic after $(k=1)$-term, the expression reduces to the
equation of the tetra-cuspi-hypocycloid, or the astroid. The fact that
the caustic for the Chang--Refsdal lens with $0\le|\gamma|<1$ has an
astroid-like shape has been widely acknowledged. In fact, the astroid
is the generic shape of the caustic for a lens system under a weak
quadrupole perturbation \citep{Ko87,An05}. Here, the $\phi$-period of
both curves are $2\pi$, and both are simply connected. Thus, the area
under each curve is
\begin{eqnarray}\lefteqn{
S[z_\mathrm c]=\sum_{k=0}^\infty
\frac{2\Gamma(k+1/2)\Gamma(k+3/2)}{\Gamma(k+1)^2}|\gamma|^{2k}
}\nonumber\\*&&=
\pi\,{}_2F_1\left(\frac12,\frac32;1;|\gamma|^2\right)
\label{eq:ccg}\\\lefteqn{
S[\zeta_\mathrm c]=\sum_{k=0}^\infty
\frac{2\Gamma(k+1/2)\Gamma(k+5/2)}{\Gamma(k+2)^2}|\gamma|^{2k+2}
}\nonumber\\*&&=2\pi\left[
1-{}_2F_1\left(-\frac12,\frac32;1;|\gamma|^2\right)\right]
\label{eq:crg}
\end{eqnarray}
where ${}_2F_1(a,b;c;x)$ is the Gaussian hypergeometric function.
The result for the area under the caustics expressed using the
elliptic integrals (see Appendix \ref{app:ell}) can be compared to the
area under the planetary caustics due to an `external' planet
\citep[see e.g.][]{Bo00}, which is basically in the same form up to
the overall scale factor ($\propto q$) once $|\gamma|\propto d^{-2}$
has been properly chosen, where $q$ and $d$ are the mass ratio and the
separation in units of Einstein ring of the planetary companion. This
is because the effect of the primary star on microlensing due to
the planet can be well approximated by the Chang--Refsdal lens
\citep{GL92,GG97l,Bo99}. This is also the case for any point mass lens
perturbed by mass lying far outside of its Einstein ring radius, as in
a wide-separation binary for instance \citep{Do99,Bo00,An05}.

For $|\gamma|> 1$, we find the Fourier-series expansion of the critical
curve as
\begin{equation}
z_\mathrm c\mathrm e^{-\mathrm i\phi_\gamma}=
\pm\frac1{|\gamma|^{1/2}}\left\{1+\sum_{k=1}^\infty
\frac{(-1)^k}{k!}\left(\frac12\right)_k
\frac{\mathrm e^{-2k\mathrm i\tilde\phi}}{|\gamma|^k}\right\}
\end{equation}
and the caustic
\begin{equation}
\zeta_\mathrm c\mathrm e^{-\mathrm i\phi_\gamma}=
\mp|\gamma|^{1/2}\left\{\tilde d_0+
\sum_{k=2}^\infty
C_k\frac{(k-1)\mathrm e^{2k\mathrm i\tilde\phi}+
k\mathrm e^{-2(k-1)\mathrm i\tilde\phi}}
{|\gamma|^k}\right\}.
\end{equation}
where $C_k$ is given by equation (\ref{eq:ck}) and
\begin{equation}
\tilde d_0=2-\frac1{|\gamma|}.
\end{equation}
Unlike the previous case, the leading terms are given by a constant.
Hence, we find that the critical curves are `centred' at
$z=\pm|\gamma|^{-1/2}\mathrm e^{\mathrm i\phi_\gamma}$, and that the
caustics are at $\zeta=\pm(|\gamma|^{-1/2}-2|\gamma|^{1/2})\mathrm
e^{\mathrm i\phi_\gamma}$. The truncation of the series for the caustic
after the first non-constant term for this case leads to the equation
of the tri-cuspi-hypocycloid, or the deltoid. In other words, as
$|\gamma|\rightarrow\infty$, the caustics reduce to a pair of
mirror-symmetric deltoids. Here, the $\phi$-period of each separate
curve is given by $\pi$ so that the area under each is found to be
\begin{eqnarray}\lefteqn{
S[z_\mathrm c]=\sum_{k=0}^\infty
\frac{\Gamma(k+3/2)^2}{\Gamma(k+1)\Gamma(k+2)}\frac1{|\gamma|^{2k+3}}
}\nonumber\\*&&
=\frac\pi{4|\gamma|^3}\,
{}_2F_1\left(\frac32,\frac32;2;\frac1{|\gamma|^2}\right)
\label{eq:ccl}\\\lefteqn{
S[\zeta_\mathrm c]=\sum_{k=0}^\infty
\frac{\Gamma(k+3/2)^2}{\Gamma(k+1)\Gamma(k+3)}\frac1{|\gamma|^{2k+3}}
}\nonumber\\*&&
=\frac\pi{8|\gamma|^3}
{}_2F_1\left(\frac32,\frac32;3;\frac1{|\gamma|^2}\right)
\label{eq:crl}
\end{eqnarray}
Here, the area derived is for each one of the curves, and therefore,
the total area under the critical curves and the caustics is twice
this result. Again, the results expressed in terms of elliptic
integrals are provided in Appendix \ref{app:ell}.

\subsection{Pre-caustics}
\label{sec:PC}

\begin{figure}
\includegraphics[width=\hsize]{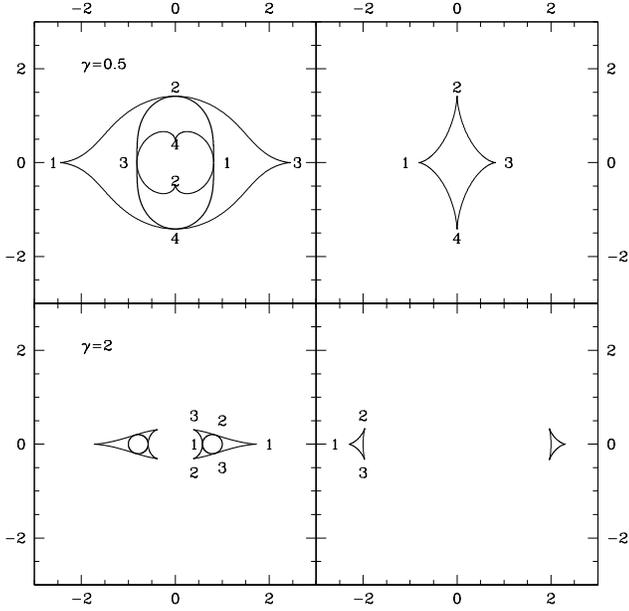}
\caption{\label{fig:crl}
Critical curves, pre-caustics, and caustics of
the Chang--Refsdal lens. The top panels are for $\gamma=1/2$ and
the bottom panels are for $\gamma=2$. The left panels show the
critical curves (thick lines) and the pre-caustics (thin lines), and
the right panels show the caustics. The numbers indicate corresponding
points mapped under the lens mapping.}
\end{figure}

Since the critical points are the degenerate zeros of the imaging
polynomial corresponding to the source on a caustic point, the
corresponding pre-caustic points can be found from the remaining zeros
of the polynomial after the two degenerate linear factors
$(z-z_\mathrm c)^2$ have been factored out. The coefficients of the
resulting quadratic polynomials $p_2z^2+p_1z+p_0$ are (for simplicity,
$\gamma$ is taken to be real)
\begin{eqnarray}
p_2&=&\gamma(1-\gamma^2)
\,;\nonumber\\
p_1&=&\mathrm e^{-\mathrm i\phi}
\left[\frac{1+\gamma\mathrm e^{2\mathrm i\phi}}
{\left(1+\gamma\mathrm e^{-2\mathrm i\phi}\right)^{1/2}}
-(1-2\gamma^2)\left(1+\gamma\mathrm e^{2\mathrm i\rm\phi}\right)^{1/2}\right]
\,;\nonumber\\
p_0&=&-\gamma\mathrm e^{-2\mathrm i\phi}(1+\gamma\mathrm e^{2\mathrm i\phi})
\end{eqnarray}
for $0\le\gamma<1$, and
\begin{eqnarray}
p_2&=&\gamma(\gamma^2-1)
\,;\nonumber\\
p_1&=&\mp\left[(2\gamma^2-1)\left(\gamma+\mathrm e^{-2i\phi}\right)^{1/2}
+\frac{1+\gamma\mathrm e^{2\mathrm i\phi}}
{\left(\gamma+\mathrm e^{2\mathrm i\phi}\right)^{1/2}}\right]
\,;\nonumber\\
p_0&=&\gamma(\gamma+\mathrm e^{-2\mathrm i\phi})
\end{eqnarray}
for $\gamma>1$. Finally, we find the parametric expression of the
pre-caustics to be
\begin{eqnarray}\lefteqn{
z_\mathrm p^+=\mathrm e^{-\mathrm i\phi}
\frac{\left(1+\gamma\mathrm e^{2\mathrm i\phi}\right)^{1/2}}
{2\gamma(1-\gamma^2)}
\left[1-2\gamma^2
-\frac{\left(1+\gamma\mathrm e^{2\mathrm i\phi}\right)^{1/2}+r}
{\left(1+\gamma\mathrm e^{-2\mathrm i\phi}\right)^{1/2}}\right]
}\label{eq:zpcp}\\\label{eq:zpcn}\lefteqn{
z_\mathrm p^-=\mathrm e^{-\mathrm i\phi}
\frac{\left(1+\gamma\mathrm e^{2\mathrm i\phi}\right)^{1/2}}
{2\gamma(1-\gamma^2)}
\left[1-2\gamma^2
-\frac{\left(1+\gamma\mathrm e^{2\mathrm i\phi}\right)^{1/2}-r}
{\left(1+\gamma\mathrm e^{-2\mathrm i\phi}\right)^{1/2}}\right]
}\end{eqnarray}
for $0\le\gamma<1$, and
\begin{equation}
z_\mathrm p=\pm\frac{\left(\gamma+\mathrm e^{-2\mathrm i\phi}\right)^{1/2}}
{2\gamma(\gamma^2-1)}
\left[2\gamma^2-1+\frac{
\left(\gamma+\mathrm e^{-2\mathrm i\phi}\right)^{1/2}
\mathrm e^{2\mathrm i\phi}+r\mathrm e^{\mathrm i\phi}}
{\left(\gamma+\mathrm e^{2\mathrm i\phi}\right)^{1/2}}\right]
\label{eq:pcp}
\end{equation}
for $\gamma>1$. Here,
\begin{equation}
\frac{r^2}2=1+\gamma\cos2\phi-(1-2\gamma^2)
\left(1+\gamma^2+2\gamma\cos2\phi\right)^{1/2}.
\end{equation}
We note that, for $0\le\gamma<1$, the pre-caustics are two separate
curves, one of which $z_\mathrm p^+(\phi)$ completely encloses the
critical curve and the other of which $z_\mathrm p^-(\phi)$ is
completely enclosed by the critical curve. We also find that
the curve $z_\mathrm p^-(\phi)$ is in the generic shape of the
bi-cuspi-epicycloid, or the nephroid whereas $z_\mathrm p^+(\phi)$,
as a whole, appears to be the inverse curve 
of $z_\mathrm p^-(\phi)$ with respect to the origin,
rotated by 90\degr\ and rescaled by a factor of $\sqrt{1-\gamma^2}$.
On the other hand, for
$\gamma>1$, the pre-caustics are two mirror-symmetric curves, each of
which encloses one of the mirror symmetric critical curves. For both
cases, the critical curves and the pre-caustics are co-tangent to one another
at the critical points corresponding to the cusp points. The examples for
$\gamma=1/2$ and $\gamma=2$ are shown in Fig.~\ref{fig:crl}.

Following \citet{Fi02}, the ratio of the areas under the caustics and
the pre-caustics provides us with the mean total magnification of the
source that lies within the caustic. Although the complicated
parametric forms of the pre-caustics indicate that the analytic
evaluation of the area under them using the present expression is
tough, following equation (\ref{eq:area}), it reduces to a
one-dimensional (complex-contour) integral, which is straightforward
to evaluate numerically. In fact, it turns out that the area under
the pre-caustics can be evaluated analytically. The result for
$\gamma>1$ is particularly notable and is given in the next section.
On the other hand, the result for $0\le\gamma<1$ is more complicated
and is simply quoted in Appendix \ref{app:areapc}.

\begin{figure}
\includegraphics[width=\hsize]{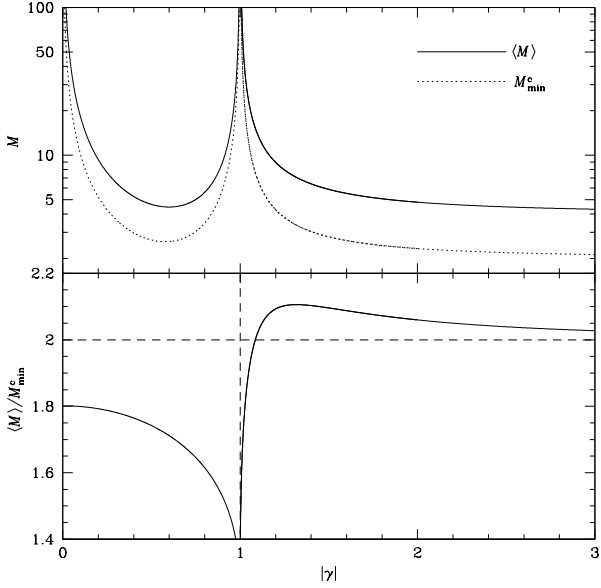}
\caption{\label{fig:mem}
Top: the mean total magnifications $\langle M
\rangle$ for the source within the caustic, that is, when there are
four images, as a function of $|\gamma|$. Also shown as a dotted line
is the minimum possible magnification $M_{\min}^\mathrm c$ for the
same configuration (that is, 4-image system). Bottom: the ratio
$\langle M\rangle/M_{\min}^\mathrm c$. The horizontal dashed line
marks the ratio of 2, which corresponds to the case that the
magnification distribution is given by a truncated power law $P\propto
M^{-3}$.}
\end{figure}

\begin{figure}
\includegraphics[width=\hsize]{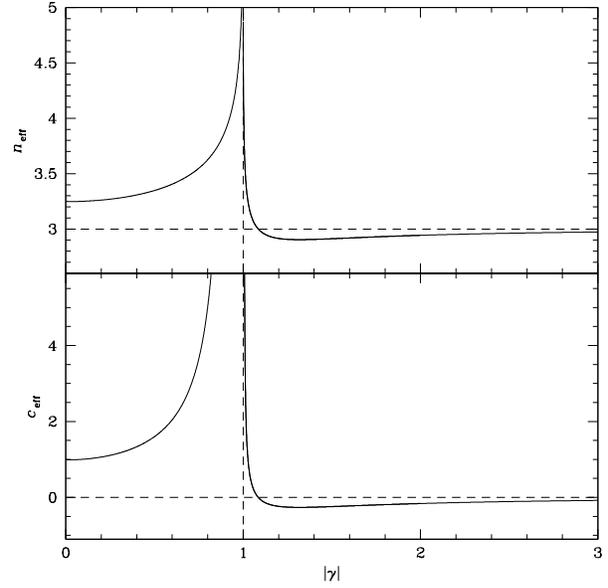}
\caption{\label{fig:eff}
Top: the effective power-index
$n_\mathrm{eff}$ for the magnification distribution. Bottom: the
effective fourth-order coefficient $c_\mathrm{eff}$ in the magnification
distribution as a function of $|\gamma|$. See the text for details.}
\end{figure}

It is known that the asymptotic form of the magnification distribution
is given by $P(M)\sim M^{-3}$ if the effect of the divergent
magnification in the vicinity of the caustics is dominant
\citep{Sc87}. If the full magnification distribution for the source
in the caustics is well approximated by its asymptotic form $P(M)\sim
M^{-3}$ down to the minimum possible value $M_{\min}^\mathrm c$, the
mean magnification is given by $\langle M\rangle=2M_{\min}^\mathrm c$.
Here, $M_{\min}^\mathrm c=[\gamma(1-\gamma^2)]^{-1}$ for $0\le\gamma<
1$ and $M_{\min}^\mathrm c=(2\gamma^2-1)/(\gamma^2-1)$ for $\gamma>1$.
In Fig.~\ref{fig:mem}, we plot $\langle M\rangle$ and $\langle
M\rangle/M_{\min}^\mathrm c$ as a function of $|\gamma|$ to examine
how much the magnification distribution deviates from its asymptotic
form. Also plotted in Fig.~\ref{fig:eff} are the effective power-index
$n_\mathrm{eff}$ for the magnification distribution defined such that
$\langle M\rangle=[(n_\mathrm{eff}-1)/(n_\mathrm{eff}-2)]
M_{\min}^\mathrm c$, which would be the mean magnification if the
magnification distribution were characterized by the power-law $P
\propto M^{-n_\mathrm{eff}}$, and the effective 4th-order coefficient
$c_\mathrm{eff}$ for the magnification distribution given from
$\langle M\rangle=[3(2+c_\mathrm{eff})/(3+2c_\mathrm{eff})]
M_{\min}^\mathrm c$, which would be the mean magnification if the
magnification distribution were characterized by $P(M)\propto M^{-3}
(1+c_\mathrm{eff}M_{\min}^\mathrm c/M)$ (note that
$1.5M_{\min}^\mathrm c<\langle M\rangle\le3M_{\min}^\mathrm c$ for
$c_\mathrm{eff}\ge-1$, that is $P(M)$ being non-negative). We find
that, for $|\gamma|\ga1.1$, $\langle M\rangle/M_{\min}^\mathrm c
\approx2$ within 5\%, implying that $P(M)\sim M^{-3}$ may indeed be a
good approximation for the magnification distribution for the source
within the deltoid-like caustics down to the minimum possible
magnification. In fact, we find the full analytic expression for the
magnification distribution for this case in section \ref{sec:LEM} and
show that it is well approximated by $P(M)\sim M^{-3}$. On the other
hand, for $0\le|\gamma|<1$, $\langle M\rangle/M_{\min}^\mathrm c$ is
smaller than 2, which implies that the power-law index should be
steeper than $-3$ if the distribution were a truncated power-law, or
that there is a significant positive higher order term if $P(M)$ is
approximated by a power-series to $M$ with the leading term as $M
\rightarrow\infty$ given by $\propto M^{-3}$. We also note that the
limiting value of $\langle M\rangle/M_{\min}^\mathrm c$ as $\gamma
\rightarrow0$ is found to be $8(3\pi)^{-1}[3\mathbf K(1/2)-2\mathbf
E(1/2)]\approx1.80149$, which translates to approximately
$c_\mathrm{eff}\approx1$. Here, $\mathbf K(x)$ and $\mathbf E(x)$ are
complete elliptic integrals of the first and second kind (see Appendix
\ref{app:ell} for the notation).

\section{Lines of Equal Individual Magnification}
\label{sec:LEM}

From equations (\ref{eq:mag}) and (\ref{eq:crisq}), it is not only the
critical curves but also the lines of equal magnification for an
individual image that can be written parametrically. That is, the
solution of
\begin{equation}
\frac1{z_m^2}-\bar\gamma=m\mathrm e^{-2\mathrm i\phi}
\end{equation}
defines the locus of image positions for which the magnification is
given by $|1-m^2|^2$. The area under each curve is proportional to the
cumulative probability that a given image is at least magnified by
$|1-m^2|^2$, and thus, its derivative with respect to $m$ is related
to the magnification distribution for a given image. Although this is
not the distribution of the total magnification for a given source
position that we want, it is still of some interest to examine this
distribution.

As before, we examine different ranges of $|\gamma|$ separately, but
now the dividing line occurs at $|\gamma|=m$ in the general case. First,
we consider the case that $0\le|\gamma|\le m$. For this case, the line
of equal magnification is
\begin{equation}
z_m=\frac{\mathrm e^{\mathrm i\phi}}{m^{1/2}}
\left(1+\frac{|\gamma|}m\mathrm e^{2\mathrm i\phi}\right)^{-1/2}
=\sum_{k=0}^\infty\frac{(-1)^k}{k!}\left(\frac12\right)_k
\frac{|\gamma|^k}{m^{k+1/2}}\mathrm e^{(2k+1)\mathrm i\phi}.
\end{equation}
This is a single simply-connected curve. The areas under it and its
derivative are
\begin{eqnarray}\lefteqn{
S[z_m]=\sum_{k=0}^\infty
\frac{2\Gamma(k+3/2)\Gamma(k+1/2)}{\Gamma(k+1)^2}
\frac{|\gamma|^{2k}}{m^{2k+1}}
}\nonumber\\*&&
=\frac\pi m\,
{}_2F_1\left(\frac12,\frac32;1;\frac{|\gamma|^2}{m^2}\right)
\label{eq:zmg}\\\lefteqn{
\frac{\mathrm d S[z_m]}{\mathrm d m}=-\sum_{k=0}^\infty
\frac{4\Gamma(k+3/2)^2}{\Gamma(k+1)^2}
\frac{|\gamma|^{2k}}{m^{2k+2}}
}\nonumber\\*&&
=-\frac\pi{m^2}\,
{}_2F_1\left(\frac32,\frac32;1;\frac{|\gamma|^2}{m^2}\right)
\label{eq:dzmg}
\end{eqnarray}
The derivative can simply be transported to the source domain by
multiplying by the Jacobian determinant (i.e., the inverse
magnification), namely
\begin{equation}
\left|\frac{\mathrm d S[\zeta_m]}{\mathrm d m}\right|
=|m^2-1|\left|\frac{\mathrm d S[z_m]}{\mathrm d m}\right|
=\frac{\pi|m^2-1|}{m^2}\,
{}_2F_1\left(\frac32,\frac32;1;\frac{|\gamma|^2}{m^2}\right).
\label{eq:prg}
\end{equation}
The `area' under the curve 
\begin{equation}
\zeta_m
=\sum_{k=0}^\infty
C_k\left[\frac{(2k-1)\mathrm e^{(2k+1)\mathrm i\phi}}{2m^{k+1/2}}
+\frac{(2k+1)\mathrm e^{-(2k-1)\mathrm i\phi}}{2m^{k-1/2}}\right]|\gamma|^k
\end{equation}
(where $C_k$ is given by equation \ref{eq:ck}), is in principle found
either by integrating this or by a similar method as in section
\ref{sec:CCC}, the result however is not necessarily the actual area
since the curve $\zeta_m$ may self-intersect, but the signed sum of
the areas of adjacent simply-connected regions.

If $|\gamma|>m\ge0$, we find that the loci of the equally magnified
images
\begin{equation}
z_m=\pm\frac1{|\gamma|^{1/2}}\left(
1+\frac m{|\gamma|}\mathrm e^{-2\mathrm i\phi}\right)^{-1/2}
=\pm\sum_{k=0}^\infty\frac{(-1)^k}{k!}\left(\frac12\right)_k
\frac{m^k}{|\gamma|^{k+1/2}}\mathrm e^{-2k\mathrm i\phi}
\end{equation}
are two separate mirror-symmetric simply-connected curves. The area
and its derivative for one of the simply-connected curves is
\begin{eqnarray}\lefteqn{
S[z_m]=\sum_{k=0}^\infty
\frac{\Gamma(k+3/2)^2}{\Gamma(k+1)\Gamma(k+2)}
\frac{m^{2k+2}}{|\gamma|^{2k+3}}
}\nonumber\\*&&
=\frac{\pi m^2}{4|\gamma|^3}\,
{}_2F_1\left(\frac32,\frac32;2;\frac{m^2}{|\gamma|^2}\right)
\label{eq:zml}\\\lefteqn{
\frac{\mathrm d S[z_m]}{\mathrm d m}=\sum_{k=0}^\infty
\frac{2\Gamma(k+3/2)^2}{\Gamma(k+1)^2}
\frac{m^{2k+1}}{|\gamma|^{2k+3}}
}\nonumber\\*&&
=\frac{\pi m}{2|\gamma|^3}\,
{}_2F_1\left(\frac32,\frac32;1;\frac{m^2}{|\gamma|^2}\right),
\label{eq:dzml}
\end{eqnarray}
whereas
\begin{equation}
\left|\frac{\mathrm d S[\zeta_m]}{\mathrm d m}\right|
=\frac{\pi m|1-m^2|}{2|\gamma|^3}\,
{}_2F_1\left(\frac32,\frac32;1;\frac{m^2}{|\gamma|^2}\right)
\label{eq:distpos}
\end{equation}
where
\begin{equation}
\zeta_m
=\mp\sum_{k=0}^\infty
C_k\frac{km^{k-1}\mathrm e^{-2(k-1)\mathrm i\phi}
+(k-1)m^k\mathrm e^{2k\mathrm i\phi}}{|\gamma|^{k-1/2}}.
\end{equation}
Now, if $|\gamma|>1>m\ge0$, equation (\ref{eq:distpos}) provides us
with the distribution of the magnification of all positive parity
image(s) for a given source position because there is at most one
positive parity image for any source position for $|\gamma|>1$. More
importantly, the total magnification accounting both positive and
negative parity images when the source lies within the deltoid
caustics is given by
\begin{equation}
M_\mathrm{tot}=\frac2{1-m^2}+\frac1{|\gamma|^2-1}\,;\qquad
\frac{\mathrm d M_\mathrm{tot}}{\mathrm d m}=\frac{4m}{(1-m^2)^2},
\end{equation}
and thus,
\begin{eqnarray}\lefteqn{
P(M_\mathrm{tot})\propto
\left|\frac{\mathrm d S[\zeta_m]}{\mathrm d m}
\frac{\mathrm d m}{\mathrm d M_\mathrm{tot}}\right|
=\frac{\pi|1-m^2|^3}{8|\gamma|^3}\,
{}_2F_1\left(\frac32,\frac32;1;\frac{m^2}{|\gamma|^2}\right)
}\nonumber\\*&&
=\frac\pi{|\gamma|^3}
\left(M_\mathrm{tot}-\frac1{|\gamma|^2-1}\right)^{-3}
{}_2F_1\left(\frac32,\frac32;1;\frac{m^2}{|\gamma|^2}\right)
\label{eq:distdel}
\end{eqnarray}
where
\begin{equation}
m^2=
1-2\left(M_\mathrm{tot}-\frac1{|\gamma|^2-1}\right)^{-1}.
\end{equation}
Since $0\le m<1$, we have
\begin{equation}
M_\mathrm{tot}\ge M_{\min}^\mathrm c=\frac{2|\gamma|^2-1}{|\gamma|^2-1}.
\label{eq:min}
\end{equation}
While equation (\ref{eq:distdel}) is an exact analytic expression for
the conditional distribution of the magnification for a source within
the deltoid caustics of the Chang--Refsdal lens with $|\gamma|> 1$, the
elliptic integrals (to which the hypergeometric function reduces; see
Appendix \ref{app:ell}) are not the most convenient functions to deal
with, and therefore, it may be better to come up with a simple and
reasonable approximation. First, we note that the leading term of the
Taylor-series expansion of equation (\ref{eq:distdel}) at $M_\mathrm{tot}=
\infty$ is found to be
\begin{equation}
\left|\frac{\mathrm d S[\zeta_m]}{\mathrm d
M_\mathrm{tot}}\right|\simeq \frac\pi{|\gamma|^3}\,
{}_2F_1\left(\frac32,\frac32;1;\frac1{|\gamma|^2}\right) \cdot
M_\mathrm{tot}^{-3}+\mathcal O(M_\mathrm{tot}^{-4}),
\end{equation}
which confirms the general result on the asymptotic form of the
magnification distribution ($\sim M^{-3}$). Next, we consider the
limiting value at $M_{\min}^\mathrm c=(2|\gamma|^2-1)/(|\gamma|^2-1)$,
that is, $\mathrm dS[\zeta_m]/\mathrm d M_\mathrm{tot}=\pi/
(2|\gamma|)^3$. We find that the coefficient for the leading term of
the Taylor series is approximately equal to $\pi[M_{\min}^\mathrm c/
(2|\gamma|)]^3$ for $\gamma\ga1.1$. In other words, the approximation
of the distribution by a simple truncated power-law
\begin{equation}
P(M_\mathrm{tot})\approx P(M_{\min}^\mathrm c)
\left(\frac{M_\mathrm{tot}}{M_{\min}^\mathrm c}\right)^{-3}
\Theta(M_\mathrm{tot}-M_{\min}^\mathrm c)
\label{eq:imp}
\end{equation}
extending down to the minimum magnification (eq.~\ref{eq:min}) is in
fact a reasonably good approximation for $\gamma\ga1.1$.

Since we have the expression for the full magnification distribution,
the mean magnification for the source within the deltoid caustics may
be derived from the normalized moment. However, the same special
property of having one positive parity image together with the
existence of the magnification invariant indicates that it can be
derived from the ratio of the areas under the caustics and the
critical curve. That is to say, for sources within the deltoid caustics,
their corresponding positive parity images completely fill the region
inside the critical curve, and thus, the ratio of the area under the
critical curve to that under the caustic
\begin{equation}
\langle M_+\rangle=\frac{2\,{}_2F_1(3/2,3/2;2;|\gamma|^{-2})}
{{}_2F_1(3/2,3/2;3;|\gamma|^{-2})}
\end{equation}
gives the mean magnification of the positive parity image. Finally,
equation (\ref{eq:maginv}) implies that the mean total magnification
$\langle M_\mathrm{tot}\rangle$ for this case is related to the mean
magnification of the positive parity image $\langle M_+\rangle$ through
\begin{equation}
\langle M_\mathrm{tot}\rangle
=2\langle M_+\rangle+\frac1{|\gamma|^2-1}
=\frac{4|\gamma|^2}{|\gamma|^2-1}
\frac{{}_2F_1(1/2,3/2;2;|\gamma|^{-2})}
{{}_2F_1(3/2,3/2;3;|\gamma|^{-2})}.
\end{equation}
Using this result, we can also derive the analytic expression for the
area under the pre-caustics in equation (\ref{eq:pcp}) as
\begin{equation}
S[z_\mathrm p]=\langle M_\mathrm{tot}\rangle S[\zeta_\mathrm c]
=\frac\pi{2|\gamma|(|\gamma|^2-1)}
{}_2F_1\left(\frac12,\frac32;2;\frac1{|\gamma|^2}\right).
\label{eq:azp}
\end{equation}

\section{A Convergent Background}
\label{sec:CB}

\begin{figure}
\includegraphics[width=\hsize]{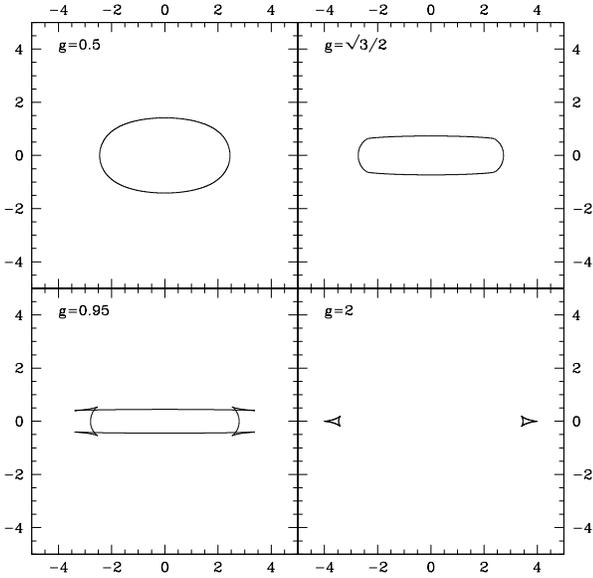}
\caption{\label{fig:conv}
Caustics for the Chang--Refsdal lens with additional
background convergence in the cases $g=1/2$, $\sqrt3/2$, $0.95$,
and $2$.}
\end{figure}

One possible generalization of the Chang--Refsdal lens equation
(\ref{eq:crlz}) is the addition of a constant convergence term
\begin{equation}
\zeta=z-\frac1{\bar z}-\gamma\bar z-\kappa z,
\end{equation}
which describes a continuous mass distribution with a constant surface
density. While the effect of the constant convergence term ($\kappa
z$) is basically an overall magnification (uniform defocusing) of the
lens domain with respect to the source domain, the specific behaviour
of the lensed images is quite distinct, depending on the sign of
$(1-\kappa)$. If $0\le\kappa<1$ (note that $\kappa$ must be
non-negative and real since Poisson's equation indicates that $\kappa$
is simply the rescaled surface mass density), introductions of new
variables $\eta=\zeta/(1-\kappa)^{1/2}$, $\xi=(1-\kappa)^{1/2}z$,
and $g=\gamma/(1-\kappa)$ simply reduce the lens equation to the form
of equation (\ref{eq:crlz}) \citep{Pa86}. Hence, the lensing behaviour
of these systems basically follows that of the standard Chang--Refsdal
lens discussed so far, except for the use of the `reduced shear' $g=
\gamma/(1-\kappa)$ and the additional focusing factor $(1-\kappa)$ in
the magnification.

However, for $\kappa>1$, which is sometimes referred to as the
over-focusing case, the introduction of the rescaled variables leaves
the sign of the point-mass deflection term changed. With $\eta=
-\zeta/(\kappa-1)^{1/2}$, $\xi=(\kappa-1)^{1/2}z$, and $g=
\gamma/(\kappa-1)$, we have the lens equation
\begin{equation}
\eta=\xi+\frac1{\bar\xi}+g\bar\xi.
\end{equation}
At the limit of $g=0$, the system allows two images
\begin{equation}
\xi=\frac\eta2\left[1\pm\left(1-\frac4{|\eta|^2}\right)^{1/2}\right]
\end{equation}
if $|\eta|\ge2$, whereas no image is formed if $|\eta|<2$. The total
magnification for $|\eta|\ge2$ is given by
\begin{equation}
(1-\kappa)^2M_\mathrm{tot}=
\frac{|\eta|^2-2}{|\eta|\sqrt{|\eta|^2-4}}
\end{equation}
where the $(1-\kappa)^2$ factor is due to the overall focusing factor.
Note that the magnification diverges at $|\eta|=2$ indicating that the
caustic of this system is given by the circle $|\eta|=2$. The
corresponding critical curve is found to be another circle $|\xi|=
1$. For this system, when the source crosses the caustic into its
interior, two images, one inside and the other outside of the critical
curve divergently merge and vanish. Two images re-appear when the
source crosses the caustics outwards.

For the general case when $g\ne0$, the lensing behaviour can be
studied using the same methods developed throughout this paper. For
example, the image positions can be found by solving similar rational
equations or polynomials as equations (\ref{eq:impol}) and
(\ref{eq:impolc}), but with the following replacement of variables:
$z\rightarrow\xi$, $\gamma\rightarrow g$, $\bar\gamma\rightarrow\bar
g$, $\zeta\rightarrow\eta$, and $\bar\zeta\rightarrow-\bar\eta$. One
notable difference between the new polynomial equation as compared to
that of the standard Chang--Refsdal lens is that it has a quadruple
zero if $g=\sqrt3/2$ and $2\eta= \pm(3+\sqrt3)\pm(3-\sqrt3)\mathrm i$,
which indicates that this system will exhibit third-order critical
behaviour, namely the swallow-tail catastrophe along the caustics, for
$g=\sqrt3/2$. In fact, this result can be more easily obtained by
analysis of the lens mapping along the lines of section \ref{sec:CCC}.

In Fig.~\ref{fig:conv}, we plot the caustics for these systems with
$g=1/2$, $\sqrt3/2$, $0.95$, and $2$. The topology of the critical
curves are basically the same as the standard null-convergent
Chang--Refsdal lens. If $0\le g<\sqrt3/2$, the caustics are the
oval-shape cuspless curve, within which no image is formed. At
$g=\sqrt3/2$, the caustic becomes a pin-cushion shape with a
swallow-tail catastrophe at its four corners. Again, the system allows
two image outside the caustics and no image inside the caustic. For
$\sqrt3/2<g<1$, the swallow-tail catastrophe of the caustics
metamorphoses to self-intersecting curves with two cusps at each of
four corners. There exist four images for the source within the
fish-tail-like region of the caustics, while there is still no image
for the source within the main pillow-like region of the caustics. As
for $g>1$, there appear two deltoid-like caustics, but unlike the
standard null-convergent case, the caustics are stretched in the
length direction rather than the width direction.

While a more detailed study of the lensing behaviour can be easily
done, it is beyond our scope here. However, we note that such a study
may help understand the distinct microlensing behaviour of the central
demagnified macro-image at the maximum of the time-delay surface
\citep[see e.g.][]{DKW05}.

\section{$\bmath{N}$ Point Masses with Shear}
\label{sec:NPM}

Another physically interesting generalization of the Chang--Refsdal
lens is to the case of $N$ point masses with shear. This can be
represented by the lens equation
\begin{equation}
\zeta=z-\sum_i\frac{m_i}{\bar z-\bar l_i}-\gamma\bar z,
\label{eq:crn}
\end{equation}
where $m_i$ are the relative masses and $l_i$ are the complex
positions. We find that the deflection function (c.f.,
eq.~\ref{eq:def})
\begin{equation}
s(z)=\sum_i^N\frac{m_i}{z-l_i}+\bar\gamma z
=\frac{\bar\gamma z\prod_i^N(z-l_i)
+\sum_i^Nm_i\prod_{j\ne i}^N(z-l_j)}{\prod_i^N(z-l_i)}
\end{equation}
is a rational function of degree $(N+1)$, where $N$ is the number of
point masses. To establish the imaging equation, we follow the same
procedure as in section \ref{sec:IP}. This leads to consideration of the
rational equation $g(z)=z-h(z)$, or equivalently the fixed points of
the mapping $h(z)=(\bar f\circ f)(z)$ where $f(z)=\bar\zeta+s(z)$ and
$\bar f(w)=\overline{f(\bar w)}$.
Here, since $f(z)$ is of degree $(N+1)$, $h(z)$ is of degree
$(N+1)^2$. In addition, as $z\rightarrow\infty$, $s(z)$ [and
consequently $f(z)$] diverges linearly ($\gamma\ne0$) and so does
$h(z)$. This indicates that the degree of the rational function
$g(z)$, as well as the order of the polynomial in its numerator, is
$(N+1)^2$ \citep{Wi90}. Obviously, as $N$ grows, finding image
positions through solving $(N+1)^2$-th order polynomial equation
quickly becomes impractical. On the other hand, the parametric
representation of the critical curves (and subsequently the caustics)
can be derived from $|s^\prime(z)|^2=1$ or by simply solving
$s^\prime(z)=-\mathrm e^{-2\mathrm i\phi}$, which reduces to a
$(2N)$-th order polynomial equation. Here, the parameter $\phi$ is
chosen such that $\phi$ is the argument angle of the tangent to the caustics
(section \ref{sec:CCC}). In other words, finding the critical curves and
the caustics is comparatively easier as the order of the relevant
polynomial grows linearly with $N$. By contrast, solving the lens
equation requires us to solve a polynomial whose order grows
quadratically as $N$ increases.

One additional point to note regarding the $N$ point mass lenses under
constant external shear is that, despite the order of the imaging
polynomial being $(N+1)^2$, the upper bound for the number of images
grows linearly with $N$. Using the result of Appendix \ref{app:no}, we
find that the total number of images for a lens system with $N$ point
masses with non-zero shear may not be greater than
$5(N+1)-6=5N-1$. [Compare this to the upper bound for $\gamma=0$,
which is $5(N-1)$ with $N\ge2$ \citep{Rh01,Rh03} and for which the
order of the corresponding imaging polynomial is $N^2+1$
\citep{Wi90}.] This also means that the imaging polynomial must have a
zero that is not a image position if $N\ge3$ (for $\gamma\ne0$ and
$|\gamma|\ne1$).  Furthermore, if $0<|\gamma|<1$, we find that
$n_+\le2N$ and $n_--n_+=N-1$ where $n_+$ and $n_-$ are the numbers of
positive and negative parity images, respectively. For $|\gamma|>1$,
these results become $n_+\le 2N-1$ and $n_--n_+=N+1$.

We note that, for the least extreme case (i.e., the source `outside'
the caustics), the difference in the number of opposite parity images
can be understood as being due to one negative parity image associated
with each point mass, plus one additional image of either positive
($0<|\gamma|<1$) or negative ($|\gamma|>1$) parity. Here, the one
additional image of `spare' parity can be regarded as the weakly
perturbed macro image: the system with $0<|\gamma|<1$ represents the
macro image at the `minimum' in the time-delay surface (the positive
parity image at sub-unity convergence), while that with $|\gamma|>1$
corresponds to the macro image at the `saddle point' in the time-delay
surface (the negative parity image).

\section{Conclusions}

This paper has presented a complete mathematical treatment of the
Chang--Refsdal lens. This simple and flexible model has found
widespread applications in astrophysics. The lens equation for the
Chang--Refsdal lens can be converted to a polynomial of the fourth
degree, the roots of which provide the image locations (together with
possibly spurious roots). If the source lies on one of the symmetry
axes with respect to the external shear, this quartic readily
factorizes into a product of quadratics, which enables the lens
equation to be solved via elementary means for these special cases.
Simple lensing invariants exist for the products of the (complex)
positions of the four images, as well as for sums of the moments of
their signed magnifications. In other words, provided the source lies
within the astroid or deltoid caustics, the sums of signed
magnification, multiplied by moments of the complex positions, are
invariant.

Of the results in this paper, we highlight just two here. First, we
have provided the equations for the pre-caustics. Only one other lens
model has been known for which the pre-caustics can be written down,
namely the singular isothermal sphere with external shear
\citep{Fi02}. Second, the exact analytical expression
(eq.~\ref{eq:distdel}) in terms of hypergeometric functions (or
elliptic integrals) has been calculated for the distribution of
magnifications for the source within the deltoid caustics of the
Chang--Refsdal lens. For many practical purposes, however, we have
shown that a truncated power-law (eq.~\ref{eq:imp}) is an excellent
approximation to this distribution.  There are a number of
applications of these two results. For example, the Chang--Refsdal lens
is often used as an approximation to the binary lens equation,
especially in the case when the lens is a star plus planet. The areas
under the critical curves and pre-caustics are directly related to the
mean magnifications of the 2 and 4 image geometries. and hence the
detectability of a planet through the gravitational microlensing
effect. Equally, the distribution of magnifications is applicable to
the statistics of high-magnification microlensing events in the
lightcurves of quasars.

\section*{acknowledgments}
The authors thank Olaf Wucknitz for a number of helpful comments.  JA
acknowledges A. O. Petters for pointing the result of \citet{Kh05},
and P. L. Schechter for that of \citet{Fi02}. Work by JA at Cambridge
(UK) was supported though the grants from the Leverhume Trust
Foundation and the Particle Physics and Astronomy Research Council
(UK). JA at MIT is supported through the grants AST-0433809 and
AST-0434277 from the National Science Foundation (US). Gravitational
lensing at Cambridge (UK) is supported by the Astrophysics Network for
Galaxy LEnsing Studies (ANGLES), funded by the European Union.

\appendix

\section{Elliptic Integrals from Hypergeometric Functions}
\label{app:ell}

We note that all hypergeometric functions found in the main text
reduce to expressions involving elliptic integrals. Although the
generality of the hypergeometric function can be seen as grounds for
superiority, there are some advantages in using elliptic integrals
especially for computational purpose, partly because numerical
routines to evaluate elliptic integrals are more readily available
than hypergeometric functions \citep[see e.g.][]{Pr92}. For this
reason, and also to help readers compare results to those found in the
literature, we provide alternative expressions written in terms of
elliptic integrals.

First, we note a relationship between hypergeometric functions with
parameters differing by unity;
\begin{displaymath}
{}_2F_1\left(a,b;c;x\right)-{}_2F_1\left(a-1,b;c;x\right)
=\frac bcx{}_2F_1\left(a,b+1;c+1;x\right),
\end{displaymath}
and one of the Euler transformations of the hypergeometric functions;
\begin{displaymath}
{}_2F_1\left(a,b;c;x\right)=
(1-x)^{c-a-b}{}_2F_1\left(c-a,c-b;c;x\right).
\end{displaymath}
Using these two, we can express any Gaussian hypergeometric
function ${}_2F_1(a,b;c;x)$ with $a$ and $b$ being half-integers
and $c$ a positive integer in terms of a linear combination of
${}_2F_1(1/2,1/2;1;x)$ and ${}_2F_1(-1/2,1/2;1;x)$ with the
coefficients being rational functions of $x$. However, the complete
elliptic integrals of the first and the second kind are actually
hypergeometric functions
\begin{eqnarray*}
\mathbf K(k)&\equiv&
\int_0^{\pi/2}\frac{\mathrm d\phi}{\sqrt{1-k^2\sin^2\phi}}
=\frac\pi2{}_2F_1\left(\frac12,\frac12;1;k^2\right)
\\
\mathbf E(k)&\equiv&
\int_0^{\pi/2}\!\mathrm d\phi\sqrt{1-k^2\sin^2\phi}
=\frac\pi2{}_2F_1\left(-\frac12,\frac12;1;k^2\right).
\end{eqnarray*}
In other words, the hypergeometric function ${}_2F_1(a,b;c;k^2)$
is in general expressible in terms of the complete elliptic integrals
if $a$ and $b$ are half-integers and $c$ is a positive integer. For
example,
\begin{eqnarray*}
\frac\pi2{}_2F_1\left(\frac12,\frac32;1;k^2\right)
&=&\frac{\mathbf E(k)}{1-k^2}
\\
\frac\pi2{}_2F_1\left(-\frac12,\frac32;1;k^2\right)
&=&2\mathbf E(k)-\mathbf K(k)
\\
\frac\pi2{}_2F_1\left(-\frac12,-\frac12;1;k^2\right)
&=&2\mathbf E(k)-(1-k^2)\mathbf K(k)
\end{eqnarray*}
etc. Further examples can be found in standard references such as
\citet{Er}, \citet{AS}, or \citet{GR}, or on-line resources like
Wolfram Research site\footnote{\url{http://functions.wolfram.com}}.
(Be careful as many authors use various different conventions for
the arguments of the elliptic integrals!)

From these, we find alternative expressions written in terms of
elliptic integrals for equations (\ref{eq:ccg}) and (\ref{eq:crg});
\begin{eqnarray}\lefteqn{
S[z_\mathrm c]=
\frac{2\mathbf E\left(|\gamma|\right)}{1-|\gamma|^2}
}\label{eq:ccg2}\\\lefteqn{
S[\zeta_\mathrm c]=
2\pi+4\mathbf K\left(|\gamma|\right)-8\mathbf E\left(|\gamma|\right),
}\label{eq:crg2}
\end{eqnarray}
for equations (\ref{eq:ccl}) and (\ref{eq:crl});
\begin{eqnarray}\lefteqn{
S[z_\mathrm c]=
\frac{|\gamma|}{|\gamma|^2-1}\mathbf E\left(\frac1{|\gamma|}\right)
-\frac1{|\gamma|}\mathbf K\left(\frac1{|\gamma|}\right)
}\label{eq:ccl2}\\\lefteqn{
S[\zeta_\mathrm c]=
\frac{2(2|\gamma|^2-1)}{|\gamma|}
\mathbf K\left(\frac1{|\gamma|}\right)
-4|\gamma|\mathbf E\left(\frac1{|\gamma|}\right),
}\label{eq:crl2}
\end{eqnarray}
for equations (\ref{eq:zmg}), (\ref{eq:dzmg}), and (\ref{eq:prg});
\begin{eqnarray}\lefteqn{
S[z_m]=
\frac{2m}{m^2-|\gamma|^2}\mathbf E\left(\frac{|\gamma|}m\right)
}\label{eq:zmg2}\\\lefteqn{
\frac{\mathrm d S[z_m]}{\mathrm d m}=
\frac2{m^2-|\gamma|^2}\left[\mathbf K\left(\frac{|\gamma|}m\right)
-\frac{2m^2}{m^2-|\gamma|^2}\mathbf E\left(\frac{|\gamma|}m\right)\right]
}\label{eq:dzmg2}\\\lefteqn{
\left|\frac{\mathrm d S[\zeta_m]}{\mathrm d m}\right|
=\frac{2|m^2-1|}{m^2-|\gamma|^2}\left[
\frac{2m^2}{m^2-|\gamma|^2}\mathbf E\left(\frac{|\gamma|}m\right)
-\mathbf K\left(\frac{|\gamma|}m\right)\right],\qquad
}\end{eqnarray}
for equations (\ref{eq:zml}), (\ref{eq:dzml}), and (\ref{eq:distpos});
\begin{eqnarray}\lefteqn{
S[z_m]=
\frac{|\gamma|}{|\gamma|^2-m^2}\mathbf E\left(\frac m{|\gamma|}\right)
-\frac1{|\gamma|}\mathbf K\left(\frac m{|\gamma|}\right)
}\label{eq:zml2}\\\lefteqn{
\frac{\mathrm d S[z_m]}{\mathrm d m}=
\frac m{|\gamma|(|\gamma|^2-m^2)}\left[
\frac{2|\gamma|^2}{|\gamma|^2-m^2}
\mathbf E\left(\frac m{|\gamma|}\right)
-\mathbf K\left(\frac m{|\gamma|}\right)\right]
}\label{eq:dzml2}\\\lefteqn{
\left|\frac{\mathrm d S[\zeta_m]}{\mathrm d m}\right|
=\frac{m|1-m^2|}{|\gamma|(|\gamma|^2-m^2)}\left[
\frac{2|\gamma|^2}{|\gamma|^2-m^2}
\mathbf E\left(\frac m{|\gamma|}\right)
-\mathbf K\left(\frac m{|\gamma|}\right)\right],\qquad}
\end{eqnarray}
and for equation (\ref{eq:azp});
\begin{equation}
S[z_\mathrm p]
=\frac{2|\gamma|}{|\gamma|^2-1}
\left[\mathbf K\left(\frac1{|\gamma|}\right)
-\mathbf E\left(\frac1{|\gamma|}\right)\right].
\label{eq:azp2}
\end{equation}

In the literature, sometimes results are given in terms of the
incomplete elliptic integrals even if they are entirely expressible
using only the complete elliptic integrals. Let us consider the Jacobi
form of the incomplete elliptic integrals of the first and the second
kind;
\begin{eqnarray*}
\mathrm F(s;k)&\equiv&
\int_0^s\frac{\mathrm d v}{(1-v^2)^{1/2}(1-k^2v^2)^{1/2}}
\\
\mathrm E(s;k)&\equiv&
\int_0^s\frac{(1-k^2v^2)^{1/2}}{(1-v^2)^{1/2}}\,\mathrm d v.
\end{eqnarray*}
Here, we note that $\mathbf K(k)=\mathrm F(1;k)$ and $\mathbf E(k)=
\mathrm E(1;k)$. Then, we find that, with $s=1/k=z^{-1}$,
\begin{eqnarray*}\lefteqn{
\mathrm F\left(\frac1z;z\right)
=\int_0^{1/z}\frac{\mathrm d v}{(1-v^2)^{1/2}(1-z^2v^2)^{1/2}}
}\\*&&
=\int_0^1\frac{\mathrm d w/z}{(1-w^2/z^2)^{1/2}(1-w^2)^{1/2}}
=\frac1z\,\mathbf K\left(\frac1z\right)
\\\lefteqn{
\mathrm E\left(\frac1z;z\right)
=\int_0^{1/z}\frac{(1-z^2v^2)^{1/2}}{(1-v^2)^{1/2}}\,\mathrm dv
=\int_0^1\frac{(1-w^2)^{1/2}}{(1-w^2/z^2)^{1/2}}\,\frac{\mathrm dw}z
}\\*&&
=\int_0^1\frac{1-z^2+z^2(1-w^2/z^2)}
{(1-w^2)^{1/2}(1-w^2/z^2)^{1/2}}\,\frac{\mathrm dw}z
\\*&&
=\frac{1-z^2}z\mathbf K\left(\frac1z\right)
+z\,\mathbf E\left(\frac1z\right)
\end{eqnarray*}
so that $z^{-1}\mathbf K(z^{-1})=\mathrm F(z^{-1};z)$ and
$z\mathbf E(z^{-1})=\mathrm E(z^{-1};z)-(1-z^2)\mathrm F(z^{-1};z)$.
From this, we also find expressions using incomplete elliptic
integrals for equations (\ref{eq:ccl2}) and (\ref{eq:crl2});
\begin{eqnarray}\lefteqn{
S[z_\mathrm c]
=\frac1{|\gamma|^2-1}\mathrm E\left(\frac1{|\gamma|};|\gamma|\right)
}\\\lefteqn{
S[\zeta_\mathrm c]
=2\mathrm F\left(\frac1{|\gamma|};|\gamma|\right)
-4\mathrm E\left(\frac1{|\gamma|};|\gamma|\right),
}\label{eq:crl3}
\end{eqnarray}
for equations (\ref{eq:zml2}) and (\ref{eq:dzml2});
\begin{eqnarray}\lefteqn{
S[z_m]
=\frac m{|\gamma|^2-m^2}
\mathrm E\left(\frac m{|\gamma|};\frac{|\gamma|}m\right)
}\\\lefteqn{
\frac{\mathrm d S[z_m]}{\mathrm d m}
=\frac1{|\gamma|^2-m^2}\left[\frac{2m^2}{|\gamma|^2-m^2}
\mathrm E\left(\frac m{|\gamma|};\frac{|\gamma|}m\right)
+\mathrm F\left(\frac m{|\gamma|};\frac{|\gamma|}m\right)\right],}
\end{eqnarray}
and for equation (\ref{eq:azp2})
\begin{equation}
S[z_\mathrm p]
=\frac2{|\gamma|^2-1}
\left[\mathrm F\left(\frac1{|\gamma|};|\gamma|\right)
-\mathrm E\left(\frac1{|\gamma|};|\gamma|\right)\right].
\label{eq:azp3}
\end{equation}

\citet{Pr92} incorporated the so-called \citet{Ca77} symmetric form of
the elliptic integrals, which may be advantageous for certain
considerations over the conventional Legendre-Jacobi standard form.
Two of Carlson's integrals are used to re-express the elliptic
integrals of the first and the second kinds:
\begin{eqnarray*}
R_F(x,y,z)&\equiv&\frac12\int_0^\infty
(t+x)^{-1/2}(t+y)^{-1/2}(t+z)^{-1/2}\,dt\\
R_D(x,y,z)&\equiv&\frac32\int_0^\infty
(t+x)^{-1/2}(t+y)^{-1/2}(t+z)^{-3/2}\,dt.
\end{eqnarray*}
From the Jacobi form of the integrals given above, by changing the
integration variable to $t=(s/v)^2-1$, it is easy to establish the
transformations:
\begin{eqnarray*}
\mathrm F(s;k)&=&sR_F(1-s^2,1-k^2s^2,1)\\
\mathrm E(s;k)&=&sR_F(1-s^2,1-k^2s^2,1)-\frac{k^2s^3}3R_D(1-s^2,1-k^2s^2,1),
\end{eqnarray*}
and subsequently, $\mathbf K(k)=R_F(0,1-k^2,1)$ and
$\mathbf E(k)=R_F(0,1-k^2,1)-(k^2/3)R_D(0,1-k^2,1)$. Using this,
we can re-express equations (\ref{eq:ccg2}) and (\ref{eq:crg2})
\begin{eqnarray*}\lefteqn{
S[z_\mathrm c]=\frac2{1-|\gamma|^2}\left[
R_F\left(0,c^2,1\right)-\frac{|\gamma|^2}3R_D\left(0,c^2,1\right)\right]
}\\\lefteqn{
S[\zeta_\mathrm c]=2\pi+\frac83|\gamma|^2R_D\left(0,c^2,1\right)
-4R_F\left(0,c^2,1\right)}
\end{eqnarray*}
where $c^2=1-|\gamma|^2$,
equations (\ref{eq:ccl2}) and (\ref{eq:crl2})
\begin{eqnarray*}\lefteqn{
S[z_\mathrm c]=\frac1{|\gamma|(|\gamma|^2-1)}\left[
R_F\left(0,q^2,1\right)
-\frac13R_D\left(0,q^2,1\right)\right]
}\\\lefteqn{
S[\zeta_\mathrm c]=\frac1{|\gamma|}\left[
\frac43R_D\left(0,q^2,1\right)
-2R_F\left(0,q^2,1\right)\right]}
\end{eqnarray*}
where $q^2=1-|\gamma|^{-2}$,
equations (\ref{eq:zmg2}) and (\ref{eq:dzmg2})
\begin{eqnarray*}\lefteqn{
S[z_m]=\frac{2m}{m^2-|\gamma|^2}\left[
R_F\left(0,c_m^2,1\right)
-\frac{|\gamma|^2}{3m^2}R_D\left(0,c_m^2,1\right)\right]
}\\\lefteqn{
\frac{\mathrm dS[z_m]}{\mathrm dm}=\frac2{(m^2-|\gamma|^2)^2}\left[
\frac23|\gamma|^2R_D\left(0,c_m^2,1\right)
-(m^2+|\gamma|^2)R_F\left(0,c_m^2,1\right)\right]}
\end{eqnarray*}
where $c_m^2=1-(|\gamma|/m)^2$,
equations (\ref{eq:zml2}) and (\ref{eq:dzml2})
\begin{eqnarray*}\lefteqn{
S[z_m]=\frac{m^2}{|\gamma|(|\gamma|^2-m^2)}\left[
R_F\left(0,q_m^2,1\right)-\frac13R_D\left(0,q_m^2,1\right)\right]
}\\\lefteqn{
\frac{\mathrm dS[z_m]}{\mathrm dm}=\frac m{|\gamma|(|\gamma|^2-m^2)^2}
\left[(|\gamma|^2+m^2)R_F\left(0,q_m^2,1\right)
-\frac23m^2R_D\left(0,q_m^2,1\right)\right]}
\end{eqnarray*}
where $q_m^2=1-(m/|\gamma|)^2$,
and equation (\ref{eq:azp2})
\begin{displaymath}
S[z_\mathrm p]=\frac2{3|\gamma|(|\gamma|^2-1)}
R_D\left(0,q^2,1\right).
\end{displaymath}

\section{The Number of Images of a Rational Lens Mapping}
\label{app:no}

In this Appendix, we examine the problem: what is the maximum number
of images for $N$ point masses with shear? In fact, it is scarcely any
more work to consider the more general case of the deflection function
that is a rational function of complex position. This problem was in
fact solved by \citet{Kh05} under still more general
considerations. In particular, for the case of $N$ point masses in the
absence of shear, their result provides us with an affirmative answer
to the conjecture by \citet[see also \citealt{Rh03}]{Rh01}: the
maximum number of images for $N$ point masses ($N\ge2$) with no shear
is $5(N-1)$. However, while the result of \citet{Kh05} is strictly
true, the limit can be further lowered depending on the particular
behaviour of the rational function, as seen by \citet{KS03}. Here, we
revisit the proofs of \citet{KS03} and \citet{Kh05}, and derive
specific limits depending on the asymptotic behaviour of the
rational lens mapping.

Let us suppose that the lens equation is written as
\begin{displaymath}
\zeta=z-\overline{s(z)}
\end{displaymath}
where $s(z)$ is a rational function of degree $r\ge2$ -- i.e, a
rational function that is neither a constant function nor a M\"obius
transformation.\footer{Formally, this indicates that the image
$s(\mathbb C_\infty)$ of the Riemann sphere under the given
(meromorphic) function $s(z):\mathbb C_\infty\rightarrow\mathbb
C_\infty$ is an $r$-fold covering of $\mathbb C_\infty$. That is, the
number of solutions in $\mathbb C_\infty$ of $s(z)=a$ (counting
multiplicities) is $r$ for any $a\in\mathbb C$. With a rational
function $s(z)=s_\mathrm n(z)/s_\mathrm d(z)$ where $s_\mathrm n(z)$
and $s_\mathrm d(z)$ are polynomials with no common factor, this is
basically $r=\max\{\deg s_\mathrm n(z),\deg s_\mathrm d(z)\}$ where
$\deg p(z)$ is the degree or the order of the polynomial $p(z)$.}
Therefore, $s(z):\mathbb C_\infty\rightarrow\mathbb C_\infty$ is
surjective -- `onto' -- but not injective -- `one-to-one'. Here
$\mathbb C_\infty\equiv\mathbb C\cup\{\infty\}$ is the `Riemann
sphere.' By considering the complex conjugate, the lens mapping can
also be written, after rearranging terms, as
\begin{displaymath}
\bar z=f(z)=\bar\zeta+s(z)
\end{displaymath}
where $f(z)$ is also a rational function of degree $r$, and in
addition $f^\prime(z)=s^\prime(z)$. Then, the image positions are
found among the fixed points of a degree-$r^2$ rational function
$h(z)=(\bar f\circ f)(z)=\bar f[f(z)]$ or equivalently the zeros of
a rational function $g(z)=z-h(z)$. Here, $\bar f(w)=\overline{f(\bar
w)}$ is also a degree-$r$ rational function of its argument so that
$h(z)$ is in fact a degree-$r^2$ rational function of $z$. On the other
hand, the degree of $g(z)$ is either $r^2+1$ if $\lim_{z\rightarrow
\infty}h(z)$ is finite or $r^2$ if it is infinite. Actually, if
$\lim_{z\rightarrow\infty}s(z)=\infty$, then $h(z)$ diverges as $z
\rightarrow\infty$, whereas $\lim_{z\rightarrow\infty}h(z)$ is finite
if $\lim_{z\rightarrow\infty}s(z)=a$ is finite except for the case
when $s(z)$ has a pole at $z=\zeta+\bar a$. Here, the incident of
the exceptional case explicitly depends on the specific choice of the
source position ($\zeta$), and so this cannot be a generic
occurrence for any lens system with an arbitrary source location.
Hence, ignoring the exceptional cases, we find that the degree of
$g(z)$ is either $r^2$ if $s(z)$ diverges as $z\rightarrow\infty$ or
$r^2+1$ if otherwise.

Now, according to the theorem due to Pierre J. L. Fatou (1878-1929)
and Gaston M. Julia (1893-1978), we have
\begin{theorem}[Fatou-Julia]
\label{FJ}
{\it ``Every attracting cycle for a rational function (degree $r\ge2$)
attracts at least one critical point,''} {\rm or equivalently}
{\it ``Every basin of attraction for a rational map (degree $r\ge2$)
contains at least one critical point.''}
\end{theorem}
The proof requires some familiarity with complex analysis, in
particular, Schwarz's Lemma and/or the Riemann Mapping Theorem, which
is beyond our scope \citep[see e.g.][]{Ca93}. Instead, here we simply
accept the implications of Theorem \ref{FJ} for the current problem as
a fact without detailing any proof: For any attracting\footer{The
fixed point of $h(z)$ is said to be super-attracting, attracting,
neutral, or repelling, if $h^\prime(z)=0$, $|h^\prime(z)|<1$,
$|h^\prime(z)|= 1$, or $|h^\prime(z)|>1$, respectively.} fixed point of $h(z)$,
there is at least one critical point\footer{The critical point of a
function is the point at which the mapping given by the function is
locally stationary. To avoid any confusion with the critical points of
the lens mapping, we will avoid the use of this term in favor of the
stationary point. For an analytic function $s(z)$, they are equivalent
to zeros of $s^\prime(z)$. However, for a meromorphic function $s(z)$
of Riemann sphere onto itself, this identification should be applied
with a caveat due to the association of $\infty^{-1}=0$. That is, the
point at which $s^\prime(z)= \infty^{-n}$ with $n$ is a positive
integer is not necessarily a critical point. For example, the
mapping $s(z)=z^{-1}$ is regular at $z=\infty$ although $s^\prime(z)=
-z^{-2}$. In fact, for the discussion in this appendix, the critical
point may be better understood as being a ramification point in
the topological sense.} (`stationary point' or `ramification point') of
$h(z)$ that converges to the fixed point under the iterative mapping
of $h(z)$, that is to say,
\begin{theorem}
\label{f2}
{\it For any point $z_\mathrm f\in\mathbb C_\infty$ that satisfies
$h(z_\mathrm f)=z_\mathrm f$ and $|h^\prime(z_\mathrm f)|<1$, there is
at least one point $z_\mathrm c\in\mathbb C_\infty$ such that
$h^\prime(z_\mathrm c)=0$ and $\lim_{n\rightarrow\infty}h^n(z_\mathrm
c)=z_\mathrm f$.}
\end{theorem}
Here, $h^2(z)=(h\circ h)(z)$, $h^3(z)=(h\circ h^2)(z)$, and so on. We
need to take particular care in the interpretation of $h^\prime(z)$
when $z=\infty$ or $h(z)=\infty$. By considering its
composition with the map $z^{-1}$, we deduce the following:
\begin{enumerate}
\item 
If $h(\infty)=\lim_{z\rightarrow\infty}h(z)=\infty$, then
$h^\prime(\infty)=\lim_{z\rightarrow\infty}z^2h^\prime(z)/[h(z)]^2$.
Specifically, if $h(z)\sim az$ as $z\rightarrow\infty$ with $a\ne0$
being a constant, then $h(\infty)=\infty$ and $|h^\prime(\infty)|=
|a|^{-1}$, and if $h(z)\sim az^n$ ($n\ge2$) as $z\rightarrow\infty$,
then $h(\infty)=\infty$ and $h^\prime(\infty)=0$.
\item
If $|h(\infty)|=|\lim_{z\rightarrow\infty}h(z)|<\infty$ is finite,
then $h^\prime(\infty)=0$ if and only if $\lim_{z\rightarrow\infty}
z^2h^\prime(z)=0$.
\item
If $h(z_\mathrm p)=\infty$ for a finite $z_\mathrm p$, then
$h^\prime(z_\mathrm p)=0$ if and only if $\lim_{z\rightarrow z_\mathrm
p}h^\prime(z)/[h(z)]^2=0$. In other words, if $z_\mathrm p$ is a
higher-order pole of $h(z)$ than a simple pole, then $h^\prime
(z_\mathrm p)=0$.
\end{enumerate}
In the following, we want to prove that Theorem \ref{f2} implies that
\begin{theorem}
\label{f3}
{\it For any point $z_0\in\mathbb C_\infty$ that satisfies $f(z_0)=\bar
z_0$ and $|s^\prime(z_0)|<1$, there is at least one point $z_\mathrm
c\in\mathbb C_\infty$ such that $s^\prime(z_\mathrm c)=0$ and
$\lim_{n\rightarrow\infty}(\bar f\circ f)^n(z_\mathrm c)=z_0$.}
\end{theorem}
To prove this, we first observe that $h(z_0)=(\bar f\circ f)(z_0)=\bar
f[f(z_0)]=\bar f(\bar z_0)=\overline{f(z_0)}=z_0$, that is, $z_0$ is a
fixed point of $h(z)$. Next, from the chain rule for the derivative of
a composite function,
\begin{equation}
h^\prime(z)=(\bar f\circ f)^\prime(z)=
\bar f^\prime[f(z)]f^\prime(z)=\bar s^\prime[f(z)]s^\prime(z),
\label{dcr}
\end{equation}
we find that $h^\prime(z_0)=\bar s^\prime[f(z_0)]s^\prime(z_0)=\bar
s^\prime(\bar z_0)s^\prime(z_0)=\overline{s^\prime(z_0)}s^\prime(z_0)
=|s^\prime(z_0)|^2$ and so that $|h^\prime(z_0)|<1$. (Here, $\bar s(z)
=\overline{s(\bar z)}$ or equivalently $\bar s(z)=\bar f(z)-\zeta$.)
Then, from
Theorem \ref{f2}, for each $z_0$, there exists at least one point
$z_\mathrm c$ such that $h^\prime(z_\mathrm c)=0$ and 
$\lim_{n\rightarrow\infty}h^n(z_\mathrm c)=\lim_{n\rightarrow\infty}
(\bar f\circ f)^n(z_\mathrm c)=z_0$. From equation (\ref{dcr}), the
$h^\prime(z_\mathrm c)=0$ indicates that we have either $\bar s^\prime
(w_\mathrm c)=0$ where $w_\mathrm c=f(z_\mathrm c)$, or $s^\prime
(z_\mathrm c)=0$. If the latter is the case, $z_\mathrm c$ is simply
the point that we are looking for. If the former, we find that
$s^\prime(\bar w_\mathrm c)=\overline{\bar s^\prime(w_\mathrm c)}=0$
and that
\begin{eqnarray*}\lefteqn{
\lim_{n\rightarrow\infty}(\bar f\circ f)^n(\bar w_\mathrm c)=
\lim_{n\rightarrow\infty}(\bar f\circ f)^n
[\overline{f(z_\mathrm c)}]=
\lim_{n\rightarrow\infty}(\bar f\circ f)^n
[\bar f(\bar z_\mathrm c)]}\\*&&=
\lim_{n\rightarrow\infty}\bar f[(f\circ\bar f)^n(\bar z_\mathrm c)]=
\bar f\left[\lim_{n\rightarrow\infty}
(f\circ\bar f)^n(\bar z_\mathrm c)\right]\\*&&=
\bar f\left[\overline{\lim_{n\rightarrow\infty}
(\bar f\circ f)^n(z_\mathrm c)}\right]=
\bar f(\bar z_0)=\overline{f(z_0)}=z_0.
\end{eqnarray*}
Here, we have exploited the associativity of the composition of
functions and the commutativity between the limit and the any analytic
map. We conclude that $\bar w_\mathrm c$ is the point that we are
looking for.  While it is not of direct interest here, following
similar arguments, we can further show that, for any point $z_0$
satisfying the condition part of Theorem \ref{f3}, there are at least
$(r+1)$ points $z_\mathrm c$ -- where $r$ is the degree of $s(z)$ --
such that $h^\prime (z_\mathrm c)=0$ and
$\lim_{n\rightarrow\infty}h^n(z_\mathrm c)=z_0$ (i.e., the consequence
part of Theorem \ref{f2}). In particular, at least one of them
satisfies $s^\prime(z_\mathrm c)=0$, while $r$ of them map to a single
point $w_\mathrm c$ under $f$ -- i.e., they are all of the $r$
pre-images of $w_\mathrm c$ -- where $w_\mathrm c$ satisfies $\bar
s^\prime(w_\mathrm c)=0$.

Note that the condition part of Theorem \ref{f3} is actually
equivalent to $z_0$ being an actual image position with $\mathcal
J(z_0)>0$. That is, $f(z_0)=\bar\zeta+s(z_0)=\bar z_0$ so that
$\zeta=z_0-\bar s(\bar z_0)$, which is just the lens equation, whereas
$\mathcal J(z_0)=1-|s^\prime(z_0)|^2=(1-|s^\prime(z_0)|)
(1+|s^\prime(z_0)|)>0$ if (and only if) $|s^\prime(z_0)|<1$. Hence,
the immediate corollary of Theorem \ref{f3} is that
\begin{corollary}[\citealt{Rh01}]
{\it The number of positive parity images ($n_+$) for the lens mapping
$\zeta=z-\overline{s(z)}$ that can be described by a rational function
$s(z)$ of degree $r\ge2$ is bounded by the number of stationary points
of $s(z)$.}
\end{corollary}
However, it is known that the latter number is exactly $2r-2$, if the
point at infinity is included and the multiplicity has been
counted,\footer{This follows from the Riemann--Hurwitz formula. That is,
since $s(z):\mathbb C_\infty\rightarrow\mathbb C_\infty$ is an
analytic map of degree $r$ of the Riemann sphere (whose Euler
characteristic is 2) onto itself, the total number of ramification
points (or stationary points) is $2r-2$. In particular, for a
polynomial $p(z)$ of order $r$, its derivative is a polynomial of
order $(r-1)$ so there are $(r-1)$ stationary points in $\mathbb C$.
In addition, since $p(z)\sim az^r$ ($a\ne0$), the point at infinity is
degenerate $(r-1)$ times, and thus it contribute an additional $(r-1)$
counts of ramification points.} and therefore, $n_+\le2(r-1)$
\citep{Rh01,Kh05}.

While this limit is strictly correct, the number of \emph{distinct}
stationary points of $s(z)$ may be smaller than $2r-2$, if there are
degenerate solutions for $s^\prime(z)=0$. In particular, if there
exists a point $z_\mathrm c$ such that $s^\prime(z_\mathrm c)=\ldots=
s^{(n-1)}(z_\mathrm c)=0$ and $s^{(n)}(z_\mathrm c)\ne0$ or if $s(z)$
possesses an $n$-th order pole, with $n\ge3$, the number of distinct
stationary points and consequently the limit on $n_+$ is lowered by
$n-2$ for each such point. Another possibility for lowering this limit
further depends on the asymptotic behaviour of $s(z)$, provided that
we are only interested in images with $|z_0|<\infty$.
\begin{itemize}
\item
First, if $\lim_{z\rightarrow\infty}s(z)=a$ is finite, $z=\infty$ is
not a fixed point of $h(z)$ provided that $s(z)$ does not have a pole
at $\zeta+\bar a$. On the other hand, if $s(z)\sim a+bz^{-n}$ where
$b\ne0$ is a constant and $n\ge3$ is an integer, then $s^\prime
(\infty)=\ldots=s^{(n-1)}(\infty)=0$ and $s^{(n)}(\infty)\ne0$ so that
$z=\infty$ is an $(n-1)$-times degenerate stationary point.
Consequently, we lose $(n-2)$ distinct points that might have
iteratively converged to one of the positive parity images, and thus,
the upper limit on $n_+$ is lowered to $n_+\le2r-n$.
\item
Secondly, if $s(z)$ diverges at least quadratically, we find not only
that $s^\prime(\infty)=h^\prime(\infty)=0$, but also that $h(\infty)=
\infty$. In other words, $z=\infty$ is a stationary point \emph{and} a
(super-attracting) fixed point. For this case, the point $z=\infty$
obviously does not iteratively converge to any finite fixed point, and
therefore, this point should not be considered for the limit for $n_+$
at all (counting all of its multiplicity). If $s(z)\sim az^n$ ($a\ne0$
and $n\ge2$), we find that $s^\prime(\infty)=\ldots=s^{(n-1)}(\infty)=
0$ and $s^{(n)}\ne0$, and thus, the limit is lowered to $n_+\le2r-n-1$.
\item
Finally, if $s(z)$ diverges linearly so that $s(z)\sim az$ ($a\ne0$),
then $s^\prime(\infty)=a^{-1}\ne0$ and $s(\infty)=f(\infty)=\infty$
and consequently, $h^\prime(\infty)=|a|^{-2}$ and $h(\infty)=\infty$.
With the identification of $\bar\infty=\infty$, we can conclude that
there is at least one finite point such that $s^\prime(z_\mathrm c)=0$
and $\lim_{n\rightarrow\infty}h^n(z_\mathrm c)=\infty$ if $|a|>1$
since $z=\infty$ is a non-stationary attracting fixed point.
Therefore, for $|a|>1$, there is one less point with $s^\prime(z)=0$
that can iteratively converge to one of the finite positive parity
images, and consequently, the limit is given by $n_+\le2r-3$. However,
if $|a|<1$, the fixed point at infinity is repelling and thus does not
affect the behaviour of the iteration of finite points, and the
original limit $n_+\le2(r-1)$ is maintained.
\end{itemize}

\subsection{Index Theorem for a rational lens mapping}

\begin{table*}
\caption{
The number of non-infinite regular images for a rational
lens mapping of degree $r\ge2$.
Here, the lens equation is given by $\zeta=z-\bar s(z)$ with $s(z)$
being a rational function of $z$ of degree $r\ge2$. The asymptotic
behaviour of $s(z)$ as $z\rightarrow\infty$ determines the maximum
possible number of non-infinite positive parity images. The positive
integer $n$ is basically the ramification index at infinity. If $s(z)$
is finite at infinity, the total count of the order of poles in
$\mathbb C$ is the same as the degree $r$ while it is $r-n$ if $s(z)$
diverges at infinity. Physically, the number of poles corresponds to
the number of point masses. The positive integer $d$ is the
duplication index, which counts the number of $m(\ge3)$-th order
poles as well as points such that $s^\prime(z_\mathrm c)=\ldots=
s^{(m-1)}(z_\mathrm c)=0$ and $s^{(m)}(z_\mathrm c)\ne0$ with
$m\ge3$. For each such point contributes $m-2$ to $d$.}
\label{tni}
\begin{tabular}{lrcr}
\hline
Asymptotic Behaviour&$n_+$&$n_--n_+$&$n_++n_-$\\
\hline
$\sim a+bz^{-n}$ ($b\ne0$ and $n\ge3$) &
$\le2r-n-d$&$r-1$&$\le5r-2n-1-2d$\\
$\sim a+bz^{-n}$ ($b\ne0$ and $n=1$, $2$) &
$\le2r-2-d$&$r-1$&$\le5r-5-2d$\\
$\sim az$ ($0<|a|<1$) &
$\le2r-2-d$&$r-2$&$\le5r-6-2d$\\
$\sim az$ ($|a|>1$) &
$\le2r-3-d$&$r$&$\le5r-6-2d$\\
$\sim az^n$ ($a\ne0$ and $n\ge2$) &
$\le2r-n-1-d$&$r$&$\le5r-2n-2-2d$\\
\hline
\end{tabular}
\end{table*}

The limit on $n_+$ can be extended to the limit on the number of total
images ($n_++n_-$) by deriving the relation between the numbers of
positive and negative parity images ($n_+$ and $n_-$, respectively).
Let us first think of a function $F(z,\bar z)=s(z)-\bar z-\bar\zeta$
and
\begin{displaymath}
\mathrm d\ln F=
\frac{\upartial_zF\mathrm dz+\upartial_{\bar z}F\mathrm d\bar z}{F}
=\frac{s^\prime(z)\mathrm dz-\mathrm d\bar z}{s(z)-\bar z-\bar\zeta}.
\end{displaymath}
Then, the contour integral
\begin{displaymath}
\mathrm i\,\Delta_{\partial C}\arg F=
\oint_{\partial C}\frac{s^\prime(z)\mathrm dz-\mathrm d\bar z}
{s(z)-\bar z-\bar\zeta}
\end{displaymath}
along the boundary of a domain $C\subset\mathbb C$ vanishes, provided
that the domain $C$ does not contain any image position (i.e., zeros
of $F$) nor poles of $s(z)$ [i.e., $s(z)\ne\bar z+\bar\zeta$ and
$|s^\prime(z)|<\infty$ for $^\forall\!z\in C$]. This follows because
the integrand is exact, so that it is closed in $C$. Furthermore, for
an arbitrary domain $C$ that may contain image positions and/or poles,
the integral can be evaluated by collapsing the contour to
infinitesimally small circles around each image position and pole.

Suppose that $z_0$ is an image position that is not on the critical
curve. Then, $s(z)\simeq\bar\zeta+\bar z_0+(z-z_0)s^\prime(z_0)+
\mathcal O(|z-z_0|^2)$ and $s^\prime(z)\simeq s^\prime(z_0)+\mathcal 
O(|z-z_0|)$, and thus, the integral along the small circle $z=z_0+
\epsilon\mathrm e^{\mathrm i\varphi}$ ($0\le\varphi<2\pi$) around $z_0$
reduces to
\begin{displaymath}
\mathrm i\,\Delta_{\partial C}\arg F=
\mathrm i\int_0^{2\pi}\!\mathrm d\varphi\,
\frac{s^\prime(z_0)\mathrm e^{\mathrm i\varphi}+
\mathrm e^{-\mathrm i\varphi}+\mathcal O(\epsilon)}
{s^\prime(z_0)\mathrm e^{\mathrm i\varphi}
-\mathrm e^{-\mathrm i\varphi}+\mathcal O(\epsilon)},
\end{displaymath}
and can be evaluated as the limiting value for $\epsilon\rightarrow0$.
That is, if $|s^\prime(z_0)|<1$,
\begin{displaymath}
\int_0^{2\pi}\!\mathrm d\varphi\,
\frac{s^\prime(z_0)\mathrm e^{\mathrm i\varphi}+\mathrm e^{-\mathrm i\varphi}}
{s^\prime(z_0)\mathrm e^{\mathrm i\varphi}-\mathrm e^{-\mathrm i\varphi}}=
-\int_0^{2\pi}\!
\mathrm d\varphi\,\frac{1+s^\prime(z_0)\mathrm e^{2\mathrm i\varphi}}
{1-s^\prime(z_0)\mathrm e^{2\mathrm i\varphi}}=-2\pi,
\end{displaymath}
whereas if $|s^\prime(z_0)|>1$,
\begin{displaymath}
\int_0^{2\pi}\!\mathrm d\varphi\,
\frac{s^\prime(z_0)\mathrm e^{\mathrm i\varphi}+\mathrm e^{-\mathrm i\varphi}}
{s^\prime(z_0)\mathrm e^{\mathrm i\varphi}-\mathrm e^{-\mathrm i\varphi}}=
\int_0^{2\pi}\!
\mathrm d\varphi\,\frac{1+[s^\prime(z_0)]^{-1}\mathrm e^{-2\mathrm i\varphi}}
{1-[s^\prime(z_0)]^{-1}\mathrm e^{-2\mathrm i\varphi}}=2\pi.
\end{displaymath}
Here, both results can be obtained by series-expanding the denominator
and using $\int_0^{2\pi}\mathrm e^{\mathrm i n\varphi}\mathrm d\varphi=
0$ where $n$ is a non-zero integer.

Similarly, for a pole $z_\mathrm p$ of $s(z)$ of order $n$, by
factoring out the diverging part, it is possible to write $s(z)=
(z-z_\mathrm p)^{-n}s_\mathrm p(z)$ where $s_\mathrm p(z_\mathrm p)$
is non-zero finite and $s^\prime(z)=-n(z-z_\mathrm p)^{-n-1}s_\mathrm
p(z)+(z-z_\mathrm p)^{-n}s^\prime_\mathrm p(z)$. Consequently, the
integral along the small circle around the pole $z=z_\mathrm p+
\epsilon\mathrm e^{\mathrm i\varphi}$ ($0\le\varphi<2\phi$) is given as
\begin{displaymath}
\mathrm i\,\Delta_{\partial C}\arg F=
\mathrm i\int_0^{2\pi}\!\!\mathrm d\varphi\,
\frac
{-ns_\mathrm p(z)+\epsilon\mathrm e^{\mathrm i\varphi}s^\prime_\mathrm p(z)+
\epsilon^{n+1}\mathrm e^{(n-1)\mathrm i\varphi}}
{s_\mathrm p(z)-\epsilon^n\mathrm e^{n\mathrm i\varphi}
(\bar\zeta+\bar z_\mathrm p)-\epsilon^{n+1}\mathrm e^{(n-1)\mathrm i\varphi}}
\end{displaymath}
the limiting value of which as $\epsilon\rightarrow0$ is $-2n\pi
\mathrm i$, assuming $\lim_{z\rightarrow z_\mathrm p}s_\mathrm
p^\prime(z)$ is finite. Hence, for any domain $C$ chosen large enough
to contain all of image positions and poles of $s(z)$ in $\mathbb C$,
we find that
\begin{equation}
\Delta_{\partial C}\arg F=2\pi(n_--n_+-n_\infty)
\end{equation}
where $n_+$ and $n_-$ are the total numbers of positive
[$|s^\prime(z_0)|<1$ so that $\mathcal J(z_0)>1$] and negative
[$|s^\prime(z_0)|>1$ so that $\mathcal J(z_0)<1$] parity images,
respectively, and $n_\infty$ is the sum of orders of all poles. This
result is in fact a particular case of the argument principle applied
on the function $F$.

However, for a large enough contour $\partial C$, the integral can
also be evaluated using a contour around the point at infinity
\begin{displaymath}
\mathrm i\,\Delta_{\partial C}\arg F=
\ointctrclockwise_{\partial C}
\frac{s^\prime(z)\mathrm dz-\mathrm d\bar z}{s(z)-\bar z-\bar\zeta}=
-\ointclockwise_{\partial C}
\frac{w^{-2}s^\prime(w^{-1})\mathrm dw-\bar w^{-2}\mathrm d\bar w}
{s(w^{-1})-\bar w^{-1}-\bar\zeta}.
\end{displaymath}
The limiting result as $w\rightarrow0$ is dependent on the asymptotic
behaviour of $s(z)$ as $z\rightarrow\infty$. If $\lim_{z\rightarrow
\infty}s(z)$ is finite, using the asymptotic expansion form $s(z)\sim
a+bz^{-n}$ where $a$ and $b\ne 0$ are constants ($a$ may be nil) and
$n$ is a positive integer, the integral reduces to
\begin{displaymath}
\mathrm i\,\Delta_{\partial C}\arg F=
\oint_{\partial C}\frac{\mathrm d\bar w}{\bar w}=
-\mathrm i\int_0^{2\pi}\mathrm d\varphi=-2\pi\mathrm i.
\end{displaymath}
On the other hand, if $s(z)$ diverges at least quadratically, that is
$s(z)\sim az^n$ with $n\ge 2$ and $a\ne0$, then
\begin{displaymath}
\mathrm i\,\Delta_{\partial C}\arg F=
n\oint_{\partial C}\frac{\mathrm dw}w=
n\mathrm i\int_0^{2\pi}\mathrm d\varphi=2n\pi\mathrm i.
\end{displaymath}
Finally, if $s(z)\sim az$ ($a\ne0$) diverges linearly, then
\begin{displaymath}
\mathrm i\,\Delta_{\partial C}\arg F=
\mathrm i\int_0^{2\pi}\mathrm d\varphi
\frac{a\mathrm e^{-\mathrm i\varphi}+\mathrm e^{\mathrm i\varphi}}
{a\mathrm e^{-\mathrm i\varphi}-\mathrm e^{\mathrm i\varphi}}=
\left\{\begin{array}{cl}
2\pi\mathrm i&\mbox{if }\ |a|>1\\-2\pi\mathrm i&\mbox{if }\ |a|<1
\end{array}\right..
\end{displaymath}
By summarizing the results, we arrive at a relation for the difference
between the numbers of positive and negative parity images
\begin{equation}
n_--n_+=\left\{\begin{array}{cl}
n_\infty-1&\mbox{if }\ \lim_{z\rightarrow\infty}|z^{-1}s(z)|<1\\
n_\infty+n=r&\mbox{if }\ \lim_{z\rightarrow\infty}|z^{-1}s(z)|>1
\end{array}\right.
\end{equation}
where $n$ is the ramification index of $s(z)$ at $z=\infty$. Since
$s(z)$ is a rational function, if $s(z)$ diverges as $z\rightarrow
\infty$, $n$ is the same as the asymptotic power index of $s(z)$,
which must be an integer for this case. Furthermore, $n_\infty$ is in
fact the same as the order of the polynomial in the denominator when
$s(z)$ is expressed as a quotient of relative prime polynomials. Since
the asymptotic power index for a rational function is basically the
difference of the orders of those polynomials, and the degree of the
rational function is the larger of two, we find that $r=n_\infty+n$ if
$s(z)$ is divergent as $z\rightarrow\infty$. This result is in fact
the extension of the so-called index theorem \citep{Bu81} to the case
when the deflection function is given by a rational function.

The fact that $n_--n_+$ is a fixed number for given $s(z)$, in
particular, being independent of $\zeta$, further implies that for a
given lens system, change of source position can only create or
destroy images of opposite parity in a pair. The upper limit on the
total number of non-infinite images can be directly found from the
limit on $n_+$ using $n_++n_-=2n_++(n_--n_+)$. The final result on the
limits of number of images for a number of rational lens maps are
summarized in Table \ref{tni}.

Comparing the upper limit on the number of images to the degree of
the imaging equation $g(z)=z-h(z)=0$, which is either $r^2$ if $s(z)$
diverges as $z\rightarrow\infty$ or $r^2+1$ if otherwise, we find that
they can be same only if $r=2$ or $r=3$ [i.e.,
$r^2-5r+6=(r-2)(r-3)=0$]. If $s(z)$ is a non-degenerate M\"obius
transformation, that is, a degree-1 rational function, explicit
calculations can establish that the imaging equation reduces to either
a linear equation if $s(z)$ is a linear polynomial, or a quadratic
equation if otherwise, with all of its solutions yielding image
locations. In other words, the imaging equation must have solutions
that are not image locations if $r\ge4$. This also indicates that the
moment sum invariants that may be derived through the similar methods as
in section \ref{sec:PMM} will be valid only if the degree of the
rational deflection function is 1, 2, or 3, since, if otherwise, there
must be spurious roots.

\onecolumn
\section{The mean magnification for the source inside the caustics}
\label{app:areapc}

As noted in section \ref{sec:PC}, the ratio of the areas under the
caustics and the pre-caustics gives the mean magnification for the
source inside the caustics. So far, we have derived the areas under
the caustics (eqs.~\ref{eq:crg} and \ref{eq:crg2}) for $0\le|\gamma|<
1$ and (eqs.~\ref{eq:crl}, \ref{eq:crl2} and \ref{eq:crl3}) for
$|\gamma|>1$, and under the pre-caustic (eqs.~\ref{eq:azp},
\ref{eq:azp2} and \ref{eq:azp3}) for $|\gamma|>1$. To complete the
discussion for all values of $|\gamma|$, we need to figure out the
area between the two pre-caustics for $0\le\gamma<1$
(eqs.~\ref{eq:zpcp} and \ref{eq:zpcn}). As it turns out, this can be
expressed in terms of the standard elliptic integrals (all four forms
are equivalent):
\begin{eqnarray*}\lefteqn{
S[z_\mathrm p^+]-S[z_\mathrm p^-]
=\frac{4|\gamma|}{c^{5/2}}
\left[(1+c)\mathbf\Pi(\mu,x)+\mathbf K(x)-2c\mathbf E(x)\right]
}\\*&&
=\frac{4|\gamma|}{c^{5/2}}
\left[\frac{|\gamma|^2(1+2c)}{1+c}\mathbf\Pi(-\lambda,x)
+c(1+2c)\mathbf K(x)-2c\mathbf E(x)\right]
\\*&&
=\frac{4|\gamma|}{c^2(1+2c)^{1/2}}
\left[4|\gamma|^2\mathbf\Pi(\nu,y)+2c(1+2c)\mathbf K(y)
-(1+2c)\mathbf E(y)\right]
\\*&&
=\frac{4|\gamma|}{c^2(1+2c)^{1/2}}
\left[4c\mathbf\Pi(-\tau,y)+2\mathbf K(y)-(1+2c)\mathbf E(y)\right],
\end{eqnarray*}
%
or in terms of the Carlson symmetric form:
\begin{eqnarray*}\lefteqn{
S[z_\mathrm p^+]-S[z_\mathrm p^-]
=\frac{4|\gamma|}{c^{5/2}}\left\{
(2-c)R_F\left(0,\frac\delta\rho,1\right)-\frac16\left[
\frac{|\gamma|^2}cR_J\left(0,\frac\delta\rho,1,\frac1\rho\right)
+\frac{4|\gamma|^2-3}{(1+2c)}
R_D\left(0,\frac\delta\rho,1\right)\right]\right\}
}\\*&&
=\frac{4|\gamma|}{c^{5/2}}\left\{
R_F\left(0,\frac\delta\rho,1\right)+\frac16\left[
\frac{(1+2c)|\gamma|^2}{(1+c)^2}R_J\left(0,\frac\delta\rho,1,\delta\right)
+\frac{3-4|\gamma|^2}{(1+2c)}
R_D\left(0,\frac\delta\rho,1\right)\right]\right\}
\\*&&
=\frac{4|\gamma|}{c^2(1+2c)^{1/2}}\left\{
3R_F\left(0,\frac\rho\delta,1\right)-\frac13\left[
\frac{4|\gamma|^2}{(1+2c)}R_J\left(0,\frac\rho\delta,1,\frac1\delta\right)
+\frac{3-4|\gamma|^2}{(1+2c)}
R_D\left(0,\frac\rho\delta,1\right)\right]\right\}
\\*&&
=\frac{4|\gamma|}{c^2(1+2c)^{1/2}}\left[
(1+2c)R_F\left(0,\frac\rho\delta,1\right)+\frac13\left[
\frac{4c|\gamma|^2}{(1+c)^2}R_J\left(0,\frac\rho\delta,1,\rho\right)
+\frac{4|\gamma|^2-3}{(1+2c)}
R_D\left(0,\frac\rho\delta,1\right)\right]\right\}.
\end{eqnarray*}
Here $c=\sqrt{1-|\gamma|^2}$,
\begin{displaymath}
\lambda=\frac1{2(1+c)}\,,\qquad
\mu=\frac{|\gamma|^2}{2c(1+c)}\,,\qquad
x^2=\frac{3-4|\gamma|^2}{4c(1+2c)}\,;
\qquad
\tau=\frac{|\gamma|^2}{(1+c)^2}\,,\qquad
\nu=\frac1{1+2c}\,,\qquad
y^2=\frac{4|\gamma|^2-3}{(1+2c)^2}\,,
\end{displaymath}
\begin{displaymath}
\rho=\frac\tau\mu=1-\tau=\frac1{1+\mu}=\frac{2c}{1+c}\,;
\qquad
\delta=\frac\lambda\nu=1-\lambda=\frac1{1+\nu}=\frac{1+2c}{2(1+c)}\,;
\qquad
1-x^2=\frac\delta\rho=\frac{1+2c}{4c}\,;\qquad
1-y^2=\frac\rho\delta=\frac{4c}{1+2c},
\end{displaymath}
and $\mathbf\Pi(n,k)$ is the complete elliptic integral of the third
kind:
\begin{displaymath}
\mathbf\Pi(n,k)\equiv\int_0^{\pi/2}
\frac{\mathrm d\phi}{\left(1+n\sin^2\phi\right)\sqrt{1-k^2\sin^2\phi}}
\end{displaymath}
and $R_J(x,y,z,p)$ is an additional Carlson's integral:
\begin{displaymath}
R_J(x,y,z,p)\equiv
\frac32\int_0^\infty(t+x)^{-1/2}(t+y)^{-1/2}(t+z)^{-1/2}(t+p)^{-1}dt.
\end{displaymath}
Note that $\mathbf\Pi(n,k)=R_F(0,1-k^2,1)-(n/3)R_J(0,1-k^2,1,1+n)$.
With these results, we can find the analytic expressions for the mean
magnification for the source inside the caustics.

As a simple illustration of the use of these expressions, we derive
the asymptotic expansions of $\langle M\rangle/M_{\min}^\mathrm c$,
which are also plotted in the lower panel of Fig.~\ref{fig:mem}:
\begin{eqnarray*}\lefteqn{
\frac{\langle M\rangle}{M_{\min}^\mathrm c}\simeq\frac8{3\pi}
\left[3\mathbf K\left(\frac12\right)-2\mathbf E\left(\frac12\right)\right]
-\frac{|\gamma|^2}{2\pi}
\left[13\mathbf K\left(\frac12\right)-14\mathbf E\left(\frac12\right)\right]
-\frac{|\gamma|^4}{288\pi}
\left[123\mathbf K\left(\frac12\right)-98\mathbf E\left(\frac12\right)\right]
+\mathcal O(|\gamma|^6),
\qquad(|\gamma|\rightarrow0)
}\\\lefteqn{
\frac{\langle M\rangle}{M_{\min}^\mathrm c}\simeq1-\frac2{4-\pi+2\ln(c/4)}
+\frac{c^2}2\left\{\frac{3\pi-8}{4-\pi+2\ln(c/4)}
+\frac{10-3\pi}{[4-\pi+2\ln(c/4)]^2}\right\}
}\\*&&\qquad+\frac{c^4}8\left\{\frac32+\frac{15\pi-41}{2[4-\pi+2\ln(c/4)]}
-\frac{267\pi-469-36\pi^2}{4[4-\pi+2\ln(c/4)]^2}
-\frac{(10-3\pi)^2}{[4-\pi+2\ln(c/4)]^3}\right\}
+\mathcal O(c^6),
\qquad(|\gamma|\rightarrow1^-)
\\\lefteqn{
\frac{\langle M\rangle}{M_{\min}^\mathrm c}\simeq1-\frac1{2+\ln(q/4)}
-\frac{q^2}4\left\{10-\frac{14}{2+\ln(q/4)}
+\frac3{[2+\ln(q/4)]^2}\right\}
}\\*&&\qquad+\frac{q^4}{128}\left\{392-\frac{724}{2+\ln(q/4)}
+\frac{357}{[2+\ln(q/4)]^2}
-\frac{72}{[2+\ln(q/4)]^3}\right\}
+\mathcal O(q^6),
\qquad(|\gamma|\rightarrow1^+)
\\\lefteqn{
\frac{\langle M\rangle}{M_{\min}^\mathrm c}\simeq2
+\frac1{4|\gamma|^2}-\frac1{64|\gamma|^4}+\mathcal O(|\gamma|^{-6})
\qquad(|\gamma|\rightarrow\infty).
}\end{eqnarray*}
Here, $c^2=1-|\gamma|^2$ and $q^2=1-|\gamma|^{-2}$. For the first two
cases ($0\le|\gamma|<1$), we have $M_{\min}^\mathrm c=
[|\gamma|(1-|\gamma|^2)]^{-1}$ while for the last two cases
($|\gamma|>1$), $M_{\min}^\mathrm c=(2|\gamma|^2-1)/(|\gamma|^2-1)$.

\bsp
\label{finish}
\end{document}